\documentclass[aps,prapplied,twocolumn,preprintnumbers,amsmath,amssymb,floatfix]{revtex4-2}
\usepackage{graphicx}
\usepackage{color} 
\usepackage{dcolumn}  
\usepackage{bm}
\usepackage[caption=false]{subfig}
\usepackage{float}
\usepackage{mathtools}
\usepackage{relsize}
\usepackage[utf8]{inputenc}
\usepackage[T1]{fontenc}
\usepackage{mathptmx}
\usepackage{etoolbox}
\usepackage{fix-cm}
\begin{document}
\title{Resilient Transmission of  Highly Structured Light via Active Convolved Illumination}
\author{Anindya Ghoshroy}
\author{James Davis}%
\author{Martin Wilkens}%
\affiliation{ 
Department of Electrical and Computer Engineering , Michigan Technological University, 1400 Townsend Dr, Houghton, MI, USA
}
\author{\c{S}ahin K. {\"O}zdemir}
\affiliation{
Department of Electrical and Computer Engineering, St. Louis University,  St. Louis, MO, USA
}
\author{Durdu \"O. G\"uney}
\email{dguney@mtu.edu.}

\affiliation{ 
Department of Electrical and Computer Engineering , Michigan Technological University, Houghton, MI, USA
}
\date{\today}
\begin{abstract}
Wave propagation through lossy and noisy systems as well as spatiotemporally dynamic random media often limits the complex information that can be reliably transmitted or retrieved. Progress in adaptive optics, wavefront shaping, machine learning, and wave propagation theories has significantly advanced physical and computational correction techniques. However, many approaches remain constrained by restrictive operating conditions, limiting their ability to meet growing application-specific demands, such as higher spatial resolution in imaging and greater data capacity in communications, particularly in highly dynamic environments. Recently, active convolved illumination (ACI) was proposed as a novel framework for preserving the spatial structure of optical modes during propagation through strongly distorting environments, such as atmospheric turbulence. ACI aims to bypass the prohibitive complexity of global distortion tracking by combining spectrally localized characterization with compensation using a correlated auxiliary source. Here, we provide a comprehensive review of ACI, detailing the formulation of correlation-injecting and auxiliary sources and the role of segmented reciprocal-space characterization and distortion compensation. We explore extensions in free-space optical communication and potential high-precision phase measurement, and numerically evaluate interferometric strategies for experimental implementation. This work highlights ACI as a potentially versatile framework for distortion-resilient light transmission, with broad interdisciplinary utility ranging from atmospheric physics and remote sensing to telecommunications, target tracking, and biomedical diagnosis, offering flexible deployment either as a standalone paradigm or hybridized with existing optical systems.   
\end{abstract}
\maketitle

\section{\label{sec:Intro}Introduction}

Robust light transmission through lossy and noisy systems or highly dynamic random media remains a fundamental challenge across diverse scientific and industrial domains, including biomedical imaging, oceanography, remote sensing, and telecommunications \cite{rotter2017light}. Here, random media refer to propagation environments whose optical characteristics, such as refractive index, absorption coefficients, or scattering properties, exhibit stochastic variations in space, time, or both. Examples include turbulent atmospheres and turbid media containing randomly distributed scatterers. Unlike ideal homogeneous media, which are characterized by spatiotemporally uniform material parameters, random media can induce cumulative, path-dependent scattering and interference \cite{savo2017observation,bertolotti2022imaging,gigan2022roadmap}. After propagation through random media, an optical mode exhibits intricate, temporally evolving speckle patterns. The severity of the resulting distortion is highly dependent on ambient channel conditions \cite{cao2022shaping}, thereby limiting the fidelity of information transmission and retrieval. 

For example, in systems involving wave propagation through a turbulent or turbid medium, random phase fluctuations, amplitude fading, and multiple scattering lead to severe degradation of spatial resolution, contrast, and signal fidelity \cite{groger2023advantages,khanom2024twists,jafari2025analysis}. In optical communication links, turbulence-induced beam wander, scintillation, and modal crosstalk result in increased bit error rates and reduced channel capacity \cite{jafari2025analysis,czerwinski2025atmospheric,beck2022wave}. In imaging and sensing applications, similar distortions scramble spatial information and obscure fine features, limiting penetration depth and achievable resolution and field of view (FOV) \cite{baykal2023optical,vali2025survey,arizono2023getting}.

Such challenges have spurred sustained efforts to control wave propagation and suppress distortion via physical and computational correction frameworks \cite{makris2017wave,hou2025controlling,mallejac2025active}. On the physical side, progress has been accelerated by the ubiquity of spatial light modulators \cite{mukherjee2019spatial,oes-2023-0026} alongside emerging light-manipulation techniques \cite{doi:10.1126/sciadv.adn2846}. Approaches such as adaptive optics \cite{rao2024astronomical,ZhangrefId0,brunner2024improvements} and wavefront shaping \cite{yu2022wavefront,cheng2023high,ding2024wavefront} compensate for phase distortions by probing or estimating the channel response and applying corrective spatial modulations. These physical strategies have demonstrated notable success in astronomy \cite{rao2024astronomical}, microscopy \cite{cheng2023high}, and \emph{in vivo} biological imaging \cite{she2025rapid}. Similarly, transmission-matrix-based techniques invert scattering perturbations by engineering the incident field to reconstruct target optical modes through diffusive media and biological tissues \cite{bertolotti2022imaging,ding2024wavefront,rumman2025optimizing}. In parallel, deep learning has emerged as a compelling computational tool capable of bypassing propagation-induced speckle when adequately representative training data are available \cite{guo2022blind,liu2024atmospheric}. However, supervised methods remain bottlenecked by a heavy reliance on large paired ground-truth datasets, which are frequently unavailable or prohibitively difficult to acquire in many physical systems \cite{LIAO2025131743,rs17233840}.
 
Despite their demonstrated effectiveness, current approaches tend to perform optimally only within narrow operating regimes or under controlled conditions \cite{li2015conjugate,yu2024high,zhang2026physical}. Techniques such as adaptive optics, wavefront shaping, and data-driven reconstruction may depend on precise channel estimation, iterative feedback, or exhaustive calibration \cite{yu2022wavefront,liu2024atmospheric}. The resulting characterization and calibration overhead can compromise robustness in rapidly varying environments, including those encountered in deep-tissue imaging and long-range turbulent optical links \cite{gigan2022roadmap}. However, such limitations reflect not a failure of the methods themselves but rather the inherent complexity and diversity of random media \cite{kim2025unlocking}. 

Concurrently, sustained progress has been demonstrated with the development of hybrid approaches that integrate complementary modalities, such as the correlative light-electron microscopy (CLEM) \cite{shafiei2025correlative}. Such integrated frameworks hold the potential to realize ambitious targets that elude standalone techniques, such as achieving noninvasive optical imaging at near-electron-microscopy resolution \cite{muller2021correlative,tanner2023labelling}. Ultimately, selectively uniting complementary physical and algorithmic processing has enabled sustained progress in medical imaging, sensing, and telecommunications \cite{gan2026high,lei2025vrdm}.

Against this backdrop, modern imaging and communication systems are tasked with delivering higher spatial resolution, greater data throughput, and reliable performance across strongly scattering, turbulent, and rapidly evolving media \cite{mididoddi2025threading,yu2022wavefront}. In these regimes, system performance is rarely bottlenecked by a total absence of correction strategies. Rather, it is constrained by practical physical limits, including restricted dynamic range, vulnerability to rapid decorrelation, and the severe overhead of real-time channel estimation \cite{popoff2010measuring,vynck2023light,ma2025transmission,cheng2023high}. In light of these challenges, active convolved illumination (ACI) \cite{ghoshroy2020loss,ghoshroy2020theory,adams2022incoherent} may offer a complementary approach towards robust light transmission under demanding propagation conditions.

The physical concept of ACI is inspired by studies on loss compensation in metamaterials \cite{sadatgol2015plasmon}, where the adverse effects of absorption are compensated through the introduction of an auxiliary source emulating a form of virtual gain \cite{kim2023loss,guan2023overcoming} in an otherwise passive medium \cite{ghoshroy2020loss}. The ACI concept has been extended to address a broad class of system imperfections, including absorption and attenuation in near-field coherent and incoherent imaging \cite{adams2016review,adams2016bringing,adams2017plasmonic,ghoshroy2020theory,adams2022incoherent,ghoshroy2018enhanced,adams2018plasmonic}, shot-noise-limited performance in incoherent far-field imaging \cite{adams2022incoherent}, and, more recently, propagation-induced distortions such as those arising from atmospheric turbulence for imaging \cite{ghoshroy2024enhancing} and remote sensing \cite{moazzam2025remote}.

\begin{figure*}[htbp]
\includegraphics[width=0.95\linewidth]{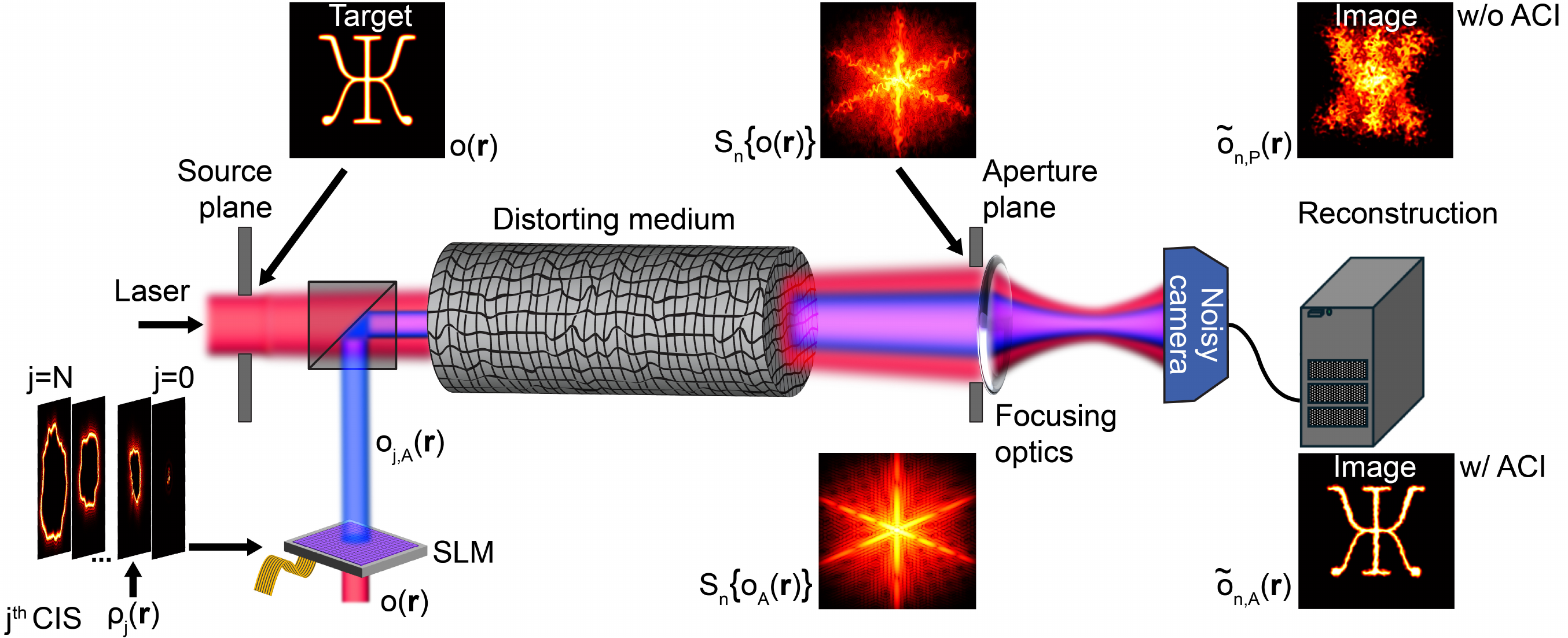}
\caption{Conceptual illustration of passive and ACI-assisted wave propagation through a random medium, using an imaging system as a representative example. The target field $o(\mathbf{r})$ propagates from the source plane through the distorting medium and a finite aperture before being focused onto a noisy camera, and reconstructed. The normalized intensity distributions above the optical path illustrate propagation without ACI, for which the accumulated distortion severely degrades the aperture-plane diffraction pattern and remains evident in the reconstructed image. In ACI-assisted propagation, the upper source port is not used. Instead, segment-specific correlation-injecting sources (CISs), $\rho_j(\mathbf{r})$, are convolved, sequentially in time, with the target field as conceptualized  by the use of a spatial light modulator (SLM) in the lower port. The intensity distributions below the optical path illustrate how this process compensates for channel-induced attenuation and spectral distortion, producing a higher-fidelity reconstruction. The subscripts $\mathrm{P}$ and $\mathrm{A}$ denote quantities obtained without and with ACI, respectively. See text for notation.}
\label{fig:ACIFullSchematic_partA}
\end{figure*}

The operating principle of ACI is schematically illustrated in Fig. \ref{fig:ACIFullSchematic_partA} using an imaging system as a representative example. The upper portion of the figure shows an example optical mode at the source plane. During propagation, the mode accumulates distortions depending on the optical inhomogeneities encountered along the propagation path shown in red. This leads to a distorted diffraction pattern at the aperture plane. Reflection, large-angle scattering, and absorption can reduce the photon flux reaching the receiver, whereas diffraction under turbulent conditions scrambles the optical field and distorts the spatial information transferred from the source plane to the aperture plane. Optical focusing followed by post-detection reconstruction of the noisy measurements may recover the general structure of the target but leaves substantial residual distortion. The achievable improvement in SNR depends in part on the severity of the channel-induced distortion, with more severely degraded signals generally requiring more sophisticated reconstruction strategies.

ACI seeks to mitigate such degradation by coherently superimposing the target field with a correlated auxiliary source \cite{sadatgol2015plasmon, ghoshroy2024enhancing}. The auxiliary source is defined as the modification to the target field and is produced during convolving the target field with a set of correlation-injecting sources (CISs), as conceptualized in the lower portion of Fig.~\ref{fig:ACIFullSchematic_partA} with the use of a spatial light modulator (SLM). Each CIS is formulated based on a spectral characterization of the uncorrected propagation system. In the idealized limit of perfect channel characterization and exact CIS implementation, the propagated auxiliary field completely compensates for the accumulated distortion and leaves no residual signature at the reconstruction plane. Consequently, the reconstructed field exactly reproduces the target distribution.   

The introduction of a CIS signifies that additional energy is supplied to the system. The term ``active'' refers exclusively to this externally supplied energy even though the propagation medium itself remains entirely passive. Furthermore, the CIS and the auxiliary source are not propagation-invariant solutions of the Helmholtz equation \cite{durmin1987exact} since the power spectral density is regulated during propagation. Consequently, the hypothesized distortion mitigation behavior should not be conflated with the behavior of eigenmodes \cite{klug2023robust,bachmann2023highly} or self-healing beams \cite{rosales2018review}. Throughout this work, the subscripts $\mathrm{P}$ and $\mathrm{A}$ are used to denote quantities associated with the passive (without ACI) and active (with ACI) configurations, respectively.   

The physical realization of the CIS depends on the propagation regime and the dominant mechanism of information degradation. For near-field imaging, hyperbolic metamaterials have been proposed as platforms for CIS generation \cite{ghoshroy2018hyperbolic,ghoshroy2018enhanced,adams2018plasmonic}. For far-field incoherent imaging, Adams \emph{et al.} demonstrated CIS generation via increased exposure through an annular aperture in the pupil plane \cite{adams2022incoherent}. In both regimes, the CIS is constructed upon the fundamental principle that shot-noise spectral variance is invariant and proportional to the total expected power across the image. More recently, a CIS formulation tailored to far-field propagation through atmospheric turbulence was introduced, leveraging a combination of stochastic spectral phase correction and selective amplification \cite{ghoshroy2024enhancing}.  

In this work, we present a comprehensive review of the ACI framework for optical information transmission through lossy and noisy systems as well as spatiotemporally dynamic random media. We provide a unified treatment of segmented reciprocal-space distortion characterization, the formulation of correlation-injecting and auxiliary sources, and the application of spectrally localized compensation through spatial-domain convolution. We examine the distinct roles of characterization and correlation injection, with particular attention to their potential advantages under strongly distorting conditions. To support experimental implementation, we numerically evaluate a receiver-side phase-shifting interferometric arrangement that reconstructs the complex probe field from intensity measurements, including the effects of detector noise on CIS estimation. We then explore extensions to free-space optical communication and potential high-precision phase measurement, examining how the preservation of structured optical fields could support information encoding and phase-sensitive readout. Finally, we outline opportunities for standalone deployment and hybridization with established optical and computational techniques, including adaptive optics, wavefront shaping, and data-driven reconstruction.

\section{\label{sec:GenericWaveProp}Wave Propagation in Random Media and Information Degradation}

We begin with a conceptual overview of wave propagation through a complex distorting medium, using an imaging system as an illustrative example. Throughout this work, we consider monochromatic waves propagating along the $z$-axis. The transverse position and transverse wavevector components are denoted by $\mathbf{r}=(x,y)$ and $\mathbf{k}=(k_x,k_y)$, respectively, with magnitudes $r=\sqrt{x^2+y^2}$  and $k=\sqrt{k_x^2+k_y^2}$. A complex field $u(\mathbf{r})$ and its reciprocal-space representation $U(\mathbf{k})$ are related through the two-dimensional Fourier transform operator $\mathcal{F}$, such that $U(\mathbf{k})=\mathcal{F}\{u(\mathbf{r})\}$.

In the following we consider the transmission and reception of a spatially coherent, monochromatic beam carrying an arbitrary transverse mode profile $o(\mathbf{r})\in\mathbb{C}$ ($\mathbf{r}\in\mathbb{R}^2$), evaluated under both passive (uncompensated) and active (ACI) configurations. These architectures are schematically illustrated in Fig.~\ref{fig:ACIFullSchematic_partA}, with representative normalized intensity distributions at key propagation planes shown in the upper and lower panels, respectively. We restrict our present analysis to transmission-mode propagation. In the passive configuration, $o(\mathbf{r})$ propagates directly into the random medium, and the beam splitter following the source plane in Fig.~\ref{fig:ACIFullSchematic_partA} is reserved exclusively for ACI integration.

Under the passive configuration, $o(\mathbf{r})$ propagates from the source plane through a spatiotemporally random medium, passes through a finite aperture, and is focused onto a receiver. Along this path, the beam suffers scattering, aberrations, and photon loss, alongside measurement noise introduced during detection. The resulting corrupted complex field distribution at the receiver is denoted by $i_n(\mathbf{r})\in\mathbb{C}$. The relationship between $i_n(\mathbf{r})$ and $o(\mathbf{r})$ can be expressed as the superposition of a propagated signal component and stochastic noise contributions arising from photon statistics and electronic noise at the detector, as \cite{walkup1974image}  
\begin{equation}
i_n(\mathbf{r}) = \mathrm{S}_n\{o(\mathbf{r})\} + N_{\mathrm{sh}}(\mathbf{r}) + N_{\mathrm{th}}(\mathbf{r}),
\label{eq:GenericWaveProp}
\end{equation}
where $\mathrm{S}_n\{\}$ is the propagation operator describing cumulative channel distortion. The term $N_{\mathrm{sh}}(\mathbf{r})$ represents signal-dependent Poisson shot noise, while $N_{\mathrm{th}}(\mathbf{r})$ denotes signal-independent Johnson--Nyquist thermal noise from the receiver electronics. Representative aperture-plane intensity distributions are depicted in the upper portion of Fig.~\ref{fig:ACIFullSchematic_partA}, prior to focusing and detection. 

The propagation operator $\mathrm{S}_n\{\}$ in Eq.~\ref{eq:GenericWaveProp} and its corresponding kernel may be deterministic or stochastic, as well as spatiotemporally invariant or dynamic. In free space, $\mathrm{S}_n\{\}$ reduces to a deterministic convolution operator governed by the coherent free-space point-spread function (PSF). In inhomogeneous media, such as turbulent atmosphere or biological tissue, the operator generalizes to a superposition integral dictated by a spatially varying, anisoplanatic PSF. Eq.~\ref{eq:GenericWaveProp} thus provides a compact, unified framework for describing both channel- and measurement-induced distortions across diverse wave-propagation regimes.

Having formulated signal degradation, we consider standard frameworks for restoring signal fidelity, which are broadly categorized by whether correction is applied before or after detection. Pre-detection compensation frameworks, including adaptive optics, wavefront shaping, and optical phase conjugation, apply corrective transformations to the optical field either before it traverses the random medium or after propagation but prior to detection. Conversely, post-detection strategies operate on the measured signal to invert channel distortions and noise. Common post-detection methods include classical deconvolution (Wiener, Richardson--Lucy, blind deconvolution), speckle-correlation frameworks relying on iterative phase retrieval \cite{gerhberg1972practical,fienup1982phase}, and data-driven machine learning models \cite{rs17233840}. While pre- and post-detection strategies alter signal and noise statistics differently, they are frequently combined within hybrid architectures to leverage complementary physical and computational advantages \cite{ji2017adaptive}. We denote the final reconstructed field as $\tilde{o}_n(\mathbf{r})$.

As imaging and communication systems push toward increasingly demanding regimes, characterized by higher spatial resolution, greater data capacity, and operation within rapidly evolving complex environments, the design of effective pre- and post-detection correction strategies becomes progressively more challenging. This fundamental difficulty arises because many conventional approaches rely on a correction derived from global characterizations of the distortion process. In practice, this necessitates retrieving a system-level representation, such as a global transfer function, speckle point-spread function (PSF), or transmission matrix (TM), to capture the aggregate action of the channel. For example, TM characterization often requires exhaustive mode scanning, whose acquisition time can exceed the millisecond-scale decorrelation time of dynamic biological media \cite{doi:10.1126/sciadv.adn2846,bertolotti2022imaging}. Similarly, speckle-correlation strategies generally rely on the spatial stationarity of the PSF, complicating their application in the wide-field, anisoplanatic environments typical of deep-tissue or long-range imaging \cite{shi2023overlapping}. Consequently, the calibration and acquisition overhead necessary to maintain global channel representations becomes prohibitively severe in dynamic, strongly scattering media \cite{gigan2022roadmap,bertolotti2022imaging}.

\section{Active Convolved Illumination: A Reciprocal-Space Framework for Enhanced Wave Propagation}

Motivated by the above limitations, the ACI framework is formulated to operate in propagation regimes where conventional correction techniques become unstable or ill-conditioned. The central premise of ACI is that the aforementioned bottlenecks may be circumvented by reformulating the correction process within reciprocal space \cite{ghoshroy2017active}. Rather than seeking a single global characterization, the framework leverages reciprocal-space segmentation, allowing the propagation-induced distortions to be characterized independently across discrete spectral segments \cite{ghoshroy2020theory}. This suppression of spectral cross-coupling was shown to enable stable correction across highly non-isoplanatic and dynamic propagation regimes \cite{adams2018plasmonic,ghoshroy2018enhanced,ghoshroy2024enhancing}. By injecting the estimated correction via physical-space convolution, ACI has been shown to enhance information fidelity for highly structured optical fields without requiring prior knowledge of their specific form.

\subsection{Reciprocal-Space Segmented Distortion Characterization and Correlation-Injecting Sources}

To formalize the reciprocal-space segmentation in the ACI framework, we first define an operational bandwidth. For the wave-propagation system in Fig.~\ref{fig:ACIFullSchematic_partA}, the operational transverse spatial-frequency bandwidth $B$ is the diameter of the circular spectral support in $\mathbf{k}$-space such that $0 \leq k \leq B/2$, where $k=|\mathbf{k}|$. The circular spectral support is partitioned into $N+1$ non-overlapping segments. A central disk ($j=0$) of diameter $W$ is centered at $k=0$, followed by $N$ concentric annular segments ($j=1,2,\ldots,N$), each having a radial width $W/2$. We define a probe beam $v_j(\mathbf{r})$ whose power spectral density (PSD) is well confined to the $j^{\mathrm{th}}$ spectral segment and negligible elsewhere. For far-field applications, such as long distance propagation through a turbulent atmosphere, $v_j(\mathbf{r})$ may be based on orbital angular momentum (OAM) modes \cite{ghoshroy2024enhancing,moazzam2025remote}, with suitably tailored Laguerre-Gaussian or Bessel-Gaussian beams. The central segment ($j=0$) may be probed using the fundamental Gaussian mode. For near-field applications, hyperbolic metamaterials with suitably tailored metal-dielectric interfaces can act as band-pass spatial filters, generating the required probe fields under localized point-source illumination \cite{ghoshroy2018hyperbolic,ghoshroy2018enhanced,adams2018plasmonic}. For the segmentation defined above, the operational bandwidth is related to the segment dimensions by $B=(N+1)W$. For a conventional far-field system, $B$ scales with the numerical aperture (NA) as $B\propto\mathrm{NA}/\lambda$ and sets the diffraction-limited resolution. In near-field systems, evanescent components may extend the accessible spectral support beyond the conventional far-field limit. In both regimes, $B$ characterizes the range of transverse spatial frequencies and the corresponding spatial-resolution scale.

\begin{figure*}[htbp]
\includegraphics[width=0.97\linewidth]{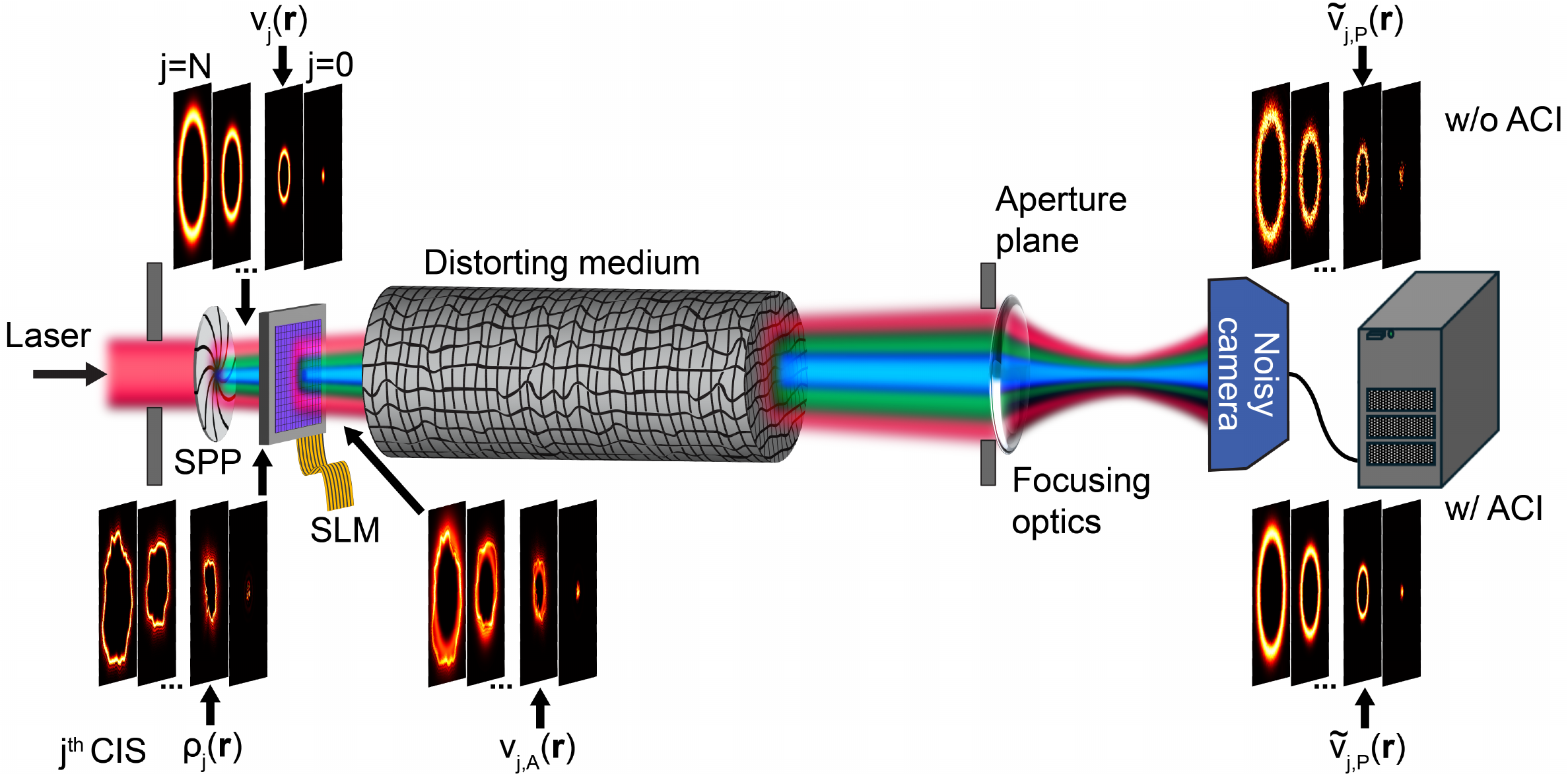}
\caption{Schematic of channel characterization and correction validation within the ACI framework. Spectrally localized probes $v_j(\mathbf{r})$ are sequentially propagated through the distorting medium, followed by optical focusing and detection. The colored regions of the diffraction cone schematically represent the propagation paths associated with different probes. Representative intensity distributions of the source probes and their reconstructions without ACI, $\tilde{v}_{j,P}(\mathbf{r})$, are shown at the upper left and upper right, respectively. Estimated corrections are encoded in CISs, $\rho_j(\mathbf{r})$, which are convolved with the corresponding probes to generate the correlation-injected probes $v_{j,A}(\mathbf{r})$. The modification $v_{j,A}(\mathbf{r})-v_j(\mathbf{r})$ constitutes the $j$\textsuperscript{th} auxiliary source. The lower portion shows representative intensity distributions of the CISs, correlation-injected probes, and reconstructed probes with ACI, $\tilde{v}_{j,A}(\mathbf{r})$, illustrating the mitigation of spatial mode distortion. SPP: spiral phase plate.}
\label{fig:ACIFullSchematic_partB}
\end{figure*}

During characterization, the ACI framework samples channel-induced distortions across the operational bandwidth using an initial trial segmentation. The distortion associated with the $j^{\mathrm{th}}$ spectral segment is probed by propagating $v_j(\mathbf{r})$ through the random medium and subsequently reconstructing the resulting field to obtain $\tilde{v}_{j,P}(\mathbf{r})$. Note that reconstruction entails compensating for the deterministic diffraction-induced phase advance associated with an ideal, distortion-free channel. Consequently, this step may be implemented either via physical optical focusing, as depicted in Fig.~\ref{fig:ACIFullSchematic_partB}, or through computational deconvolution using the known PSF of the ideal, distortion-free channel. No correction for the distortion introduced by the random medium is applied at this stage. The residual distortion in $\tilde{v}_{j,P}(\mathbf{r})$ is retained for subsequent characterization. In Fig.~\ref{fig:ACIFullSchematic_partB}, the colored regions inside the propagation cone depict the trajectories of representative probe beams, while the normalized intensity distributions for $v_j(\mathbf{r})$ and $\tilde{v}_{j,P}(\mathbf{r})$ are shown in the upper-left and upper-right panels, respectively. The deviation of the reconstructed mode profile $\tilde{v}_{j,P}(\mathbf{r})$ from the reference $v_j(\mathbf{r})$ provides a spatiotemporal snapshot of the channel-induced distortion accumulated by the $j^{\mathrm{th}}$ probe along its associated propagation path within the diffraction cone. Repeating the procedure sequentially for all $N+1$ probe beams therefore extends the characterization across the full operational bandwidth $B$.

The characterization process may be viewed as a one-time calibration, typically performed before physical deployment. The objective is to establish a spectral distortion compensation function $\mathcal{P}_j(\mathbf{k}) = \mathcal{F}\{ \rho_j(\mathbf{r})\}$, which may be expressed as
\begin{equation}
\mathcal{P}_j(\mathbf{k}) = f\left( V_j(\mathbf{k}), \tilde{V}_{j,P}(\mathbf{k}) \right), 
\label{eq:CISMapping}
\end{equation}
where $f$ denotes a functional mapping to be determined for the $j^{\mathrm{th}}$ spectral segment. Probe parameters, including peak amplitude, annular profile, phase, radial center frequency, and spectral bandwidth, are systematically varied to assess their influence on the observed spectral distortion in $\tilde{V}_{j,P}(\mathbf{k})$ \cite{ghoshroy2024enhancing,ghoshroy2018hyperbolic,ghoshroy2018enhanced}. The resulting analysis guides the formulation of $f$ in Eq.~\ref{eq:CISMapping}. Establishing $f$ is conceptually analogous to selecting and calibrating a reconstruction or correction method, such as regularized Richardson--Lucy deconvolution, a deep-learning reconstruction model, or an adaptive-optics wavefront reconstructor. The resulting compensation is intended to improve the fidelity and SNR of the reconstructed signal while limiting noise amplification and additional distortion.

\begin{figure*}[htbp]
\includegraphics[width=0.97\linewidth]{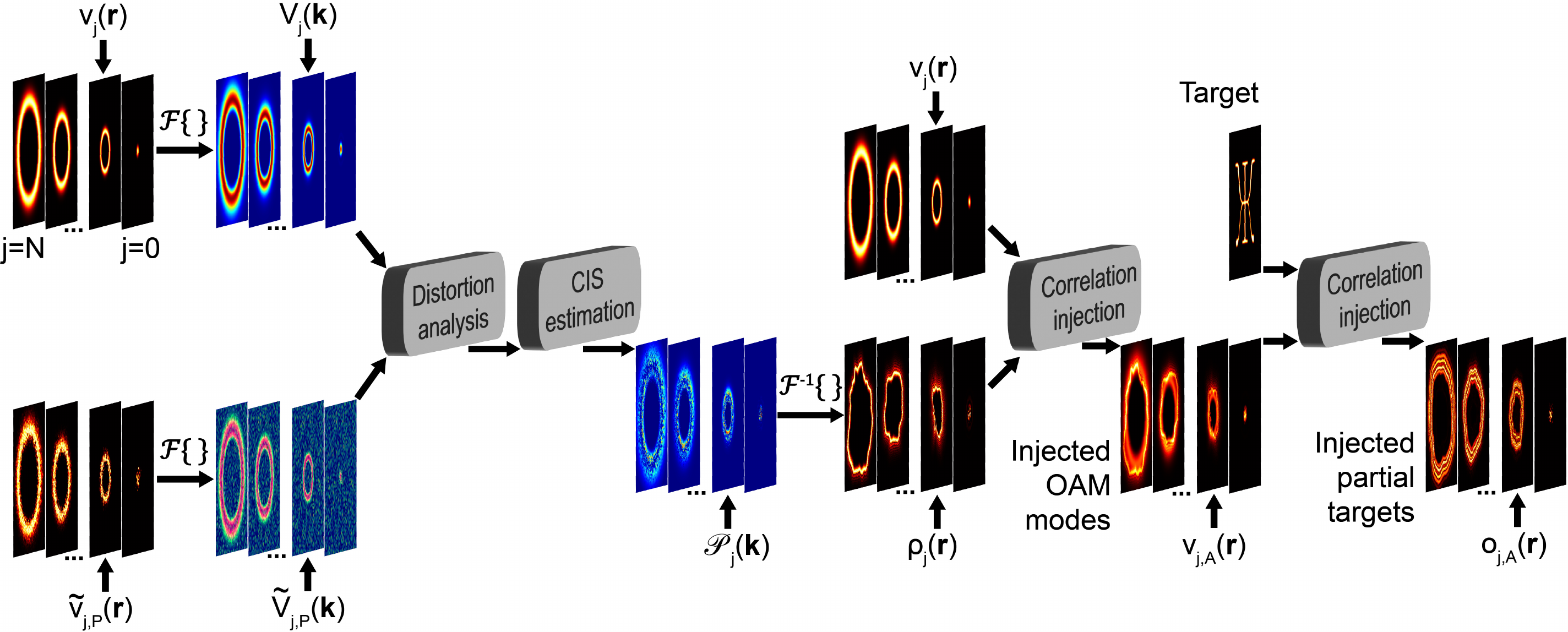}
\caption{Schematic of reciprocal-space distortion characterization, CIS estimation, and correlation injection within the ACI framework. The reference probes $v_j(\mathbf{r})$ and their reconstructed counterparts without ACI, $\tilde{v}_{j,P}(\mathbf{r})$, are Fourier transformed to obtain $V_j(\mathbf{k})$ and $\tilde{V}_{j,P}(\mathbf{k})$, respectively. The illustrated reconstructed spectra contain perturbations extending beyond the narrow spectral support of the reference probes. Spectral distortion analysis guides the estimation of the compensation functions $\mathcal{P}_j(\mathbf{k})$, whose inverse Fourier transforms define the correlation-injecting sources (CISs), $\rho_j(\mathbf{r})$. Each CIS is convolved with its corresponding reference probe to synthesize the correlation-injected probe $v_{j,A}(\mathbf{r})$. The resulting probes are subsequently used for correlation injection into the target field, generating the corresponding injected partial targets $o_{j,A}(\mathbf{r})$.
}
\label{fig:ACIFullSchematic_partC}
\end{figure*}

Within the ACI framework, sequential characterization of individual spectral segments guides the formulation of $f$ and the design of spectrally localized compensation. Previous near- and far-field implementations illustrate the potential benefits of spectral localization. In near-field superresolution imaging, compensation for absorption losses involved selective amplification of attenuated spatial-frequency components \cite{ghoshroy2018hyperbolic,ghoshroy2018enhanced,adams2018plasmonic}. Note that simply increasing the overall beam intensity does not ensure a proportional improvement in reconstruction fidelity or SNR, since signal-dependent noise also increases and bright regions may saturate detector pixels \cite{adams2022incoherent}. For far-field propagation through a turbulent medium, approximate compensation for geometric distortion in $\tilde{v}_{j,P}(\mathbf{r})$ was constructed by applying an opposing spectral phase within the corresponding bandwidth \cite{ghoshroy2024enhancing,moazzam2025remote}. Estimating a phase-correction function for the entire distorted field can become challenging when the speckle pattern is highly complex, particularly in the presence of substantial higher-order aberrations, strong scintillation, and a high density of phase branch points \cite{beck2022wave,potier2026performance}. 

Sequential characterization of the individual spectral segments yields an ensemble of $N+1$ estimated compensation functions, expressed as
\begin{equation} 
\mathcal{P}=\bigg \{\mathcal{P} _{j}(\mathbf{k})\bigg\}_{j=0}^{N}, 
\label{eq:RhoSet} 
\end{equation}
where the functions are localized to their respective spectral segments and collectively cover the full operational bandwidth $B$. The estimated compensation functions in Eq.~\ref{eq:RhoSet} are then validated through the following procedure. Structured beams with spatial-domain mode profiles $\rho_j(\mathbf{r}) = \mathcal{F}^{-1}\{ \mathcal{P}_j(\mathbf{k})\}$ are first generated. Within the ACI framework, these beams are referred to as correlation-injecting sources (CISs), and serve as the primary carriers for delivering the estimated distortion compensation to the target beam before it propagates through the distorting medium. The $j^\mathrm{th}$ probe beam is convolved with its corresponding CIS to synthesize the correlation-injected probe $v_{j,A}(\mathbf{r})$ according to
\begin{equation}
v_{j,A}(\mathbf{r}) = v_j(\mathbf{r}) \ast \rho_j(\mathbf{r}),  
\label{eq:InjectedProbe}
\end{equation}
where $\ast$ denotes the two-dimensional spatial convolution operator. The modification introduced to the original probe through Eq.~\ref{eq:InjectedProbe} is defined as the $j$\textsuperscript{th} auxiliary source and is $v_{j,A}(\mathbf{r}) - v_j(\mathbf{r})$. 

To validate the estimated compensation functions in Eq.~\ref{eq:RhoSet}, the correlation-injected probes are propagated sequentially through the medium and reconstructed to obtain the complex fields $\tilde{v}_{j,A}(\mathbf{r})$. For a spatial-domain comparison, the intensity or field-magnitude distributions of $v_j(\mathbf{r})$ and $\tilde{v}_{j,A}(\mathbf{r})$ may be compared using metrics such as normalized cross-correlation (NCC), structural similarity index measure (SSIM), and peak signal-to-noise ratio (PSNR) \cite{ghoshroy2024enhancing,moazzam2025remote}. Alternatively, validation may be performed with a reciprocal space SNR evaluated using the power spectral densities of $V_j(\mathbf{k})$ and $\tilde{V}_{j,A}(\mathbf{k})$ \cite{ghoshroy2018hyperbolic,ghoshroy2018enhanced,adams2018plasmonic}. The same metric--based evaluation is performed between $v_j(\mathbf{r})$ and $\tilde{v}_{j,P}(\mathbf{r})$. During calibration, the functional mapping $f$ in Eq.~\ref{eq:RhoSet} is refined until the selected metrics indicate the desired improvement in $\tilde{v}_{j,A}(\mathbf{r})$ relative to $\tilde{v}_{j,P}(\mathbf{r})$. The above concept is schematically illustrated in Fig.~\ref{fig:ACIFullSchematic_partB}, where representative intensity distributions of the CISs, correlation-injected probes, and corresponding reconstructed probes are shown in the lower portion. 

Previous near- and far-field studies of the ACI framework, encompassing coherent and incoherent implementations, focused on preserving intricate spatial mode profiles during propagation. Accordingly, CIS validation assessed agreement between the reconstructed and reference field-magnitude or intensity distributions. In principle, the validation criteria could be extended to assess the preservation of other beam properties, such as phase or polarization, depending on the intended application. For simplicity, the present review retains spatial field-magnitude or intensity fidelity as the basis for CIS validation, leaving the assessment of phase and polarization preservation for future work.

\subsection{Rationale for Segmented Reciprocal-Space Characterization}

It is instructive to examine the role of spectral segmentation \cite{ghoshroy2018hyperbolic,ghoshroy2018enhanced,adams2018plasmonic} both in estimating a suitable functional mapping $f$ in Eq.~\ref{eq:CISMapping} and in injecting the corresponding compensation according to Eq.~\ref{eq:InjectedProbe}. First, consider a unitary integral transform $\mathcal{T}$, such as the Fourier transform, acting on square-integrable optical fields. The operator satisfies $\mathcal{T}^{-1} = \mathcal{T}^{\dagger}$ and preserves inner products and the squared field norm, in accordance with Parseval's theorem. With appropriate normalization, the squared norm represents the total energy or power, which remains unchanged by the transform operator $\mathcal{T}$. Next, we examine the practical difficulty of characterizing strong channel-induced distortion in physical space with sufficient accuracy to construct an effective correction. We use wave propagation through a turbulent medium as a representative example. Let $v(\mathbf{r})$ denote the reference probe field and $\tilde{v}(\mathbf{r})$ its reconstructed counterpart after propagation through the medium. Under short-exposure conditions, the reconstructed field magnitude $|\tilde{v}(\mathbf{r})|$ can exhibit a speckle pattern localized around the probe mode. As turbulence strength increases, the pattern can become increasingly irregular, with more pronounced fine spatial structure \cite{cox2020structured}. More generally, estimating and implementing a complete correction in strongly scattering media can become impractical because of the large number of scattered modes, limited controllable degrees of freedom, and extensive measurement requirements. Developing suitable probe fields and measurement strategies that facilitate correction under these constraints has therefore attracted considerable research interest \cite{bertolotti2022imaging}. 

For the spectrally localized probes used in ACI, channel-induced perturbations in reciprocal space can extend across a broad range of spatial-frequency modes within the operational bandwidth $B$, as illustrated by the reconstructed probe spectrum $\tilde{V}_{j,P}(\mathbf{k})$ in the lower-left portion of Fig.~\ref{fig:ACIFullSchematic_partC}. In contrast, the reference probe spectrum $V_{j}(\mathbf{k})$, shown in the upper-left portion of the figure, is deliberately localized within the narrow band assigned to the $j^{\mathrm{th}}$ spectral segment. Because the reference probe has negligible spectral content outside its assigned band, the out-of-band components of $\tilde{V}_{j,P}(\mathbf{k})$ may be excluded during characterization when estimating $\rho _{j}(\mathbf{k})$. The $j^{\mathrm{th}}$ CIS can therefore be formulated without requiring analysis of the full distortion spectrum across the operational bandwidth $B$. Spectrally segmented characterization may partially alleviate the practical burden of correction estimation \cite{bertolotti2022imaging,gigan2022roadmap}, particularly under strongly distorting conditions, by allowing effective compensation to be determined separately within restricted spectral bands. %Therefore, a spectrally segmented characterization may partially alleviate the practical burden of correction estimation by allowing effective compensation to be determined separately within restricted spectral bands. 

It has been shown theoretically \cite{ghoshroy2024enhancing,ghoshroy2018enhanced,adams2018plasmonic} and experimentally \cite{adams2022incoherent} that, within the ACI framework, reciprocal-space redistribution of the perturbations, together with spectrally localized probing, may enable segmented distortion characterization to yield more accurate correction estimates for the $j^\mathrm{th}$ segment \cite{ghoshroy2024enhancing}. The approach may be particularly advantageous when severe distortion hinders the estimation of a single global correction function from spatial mode profiles. The benefits of spectrally localized compensation have also been demonstrated experimentally in an incoherent ACI implementation, with improvements in spectral SNR, image resolution, contrast, and resistance to pixel saturation \cite{adams2022incoherent}. Similarly, the benefits of reciprocal-space redistribution may also help retain CIS effectiveness when the compensation is injected through spatial-domain convolution. For example, numerical studies of far-field propagation through atmospheric turbulence reported relatively stable ACI performance as the refractive-index structure parameter $C_n^2$ increased from $0.25 \times 10^{-14}\,\mathrm{m}^{-2/3}$ to $3.5 \times 10^{-14}\,\mathrm{m}^{-2/3}$ \cite{ghoshroy2024enhancing}. In near-field superlens imaging simulations, successful object reconstruction was demonstrated over a finite range of progressively increasing noise variance \cite{adams2018plasmonic}. An analytical derivation of the spectral SNR enhancement in coherent near-field imaging is provided in Ref.~\cite{ghoshroy2020theory} supporting the above discussion.

During characterization, sequential segmented probing facilitates correction estimation by allowing each measured response to be analyzed independently of probes assigned to other input bands. The medium may still redistribute the $j^\mathrm{th}$ probe energy across spectral segments \cite{paterson2005atmospheric}, but the resulting response can still be attributed solely to that probe. During compensation, coherent interference with the auxiliary source shapes the injected field's spectral amplitude and phase to counteract distortion within the assigned band. The compensation does not eliminate the medium's redistribution of energy. If auxiliary sources assigned to other input bands propagate simultaneously, their energy can also be redistributed into the selected segment, where it could interfere with and degrade the compensated response. Sequential propagation of the correlation-injected probes excludes these additional contributions, isolating the compensated response associated with the $j^\mathrm{th}$ input segment from interference by other probes. 

A possible alternative to temporal isolation would be to use probes satisfying $\langle V_i(\mathbf{k}),V_j(\mathbf{k})\rangle =0$ at the source and $\langle \tilde{V}_{i,P}(\mathbf{k}),\tilde{V}_{j,P}(\mathbf{k})\rangle = 0$ after propagation and reconstruction, for $i\ne j$. Provided that the individual output responses can be resolved, such probes could enable simultaneous (i.e., parallel) characterization of the channel with multiple probes. Transmission eigenchannels in static random media offer a relevant precedent, as orthogonal input singular modes map onto mutually orthogonal output modes for a given channel realization \cite{rotter2017light}. Related studies have identified highly transmitting singular modes with reduced crosstalk over finite intervals in dynamic atmospheric turbulence \cite{bachmann2023highly}. Whether such modes can also satisfy the spectral localization required by the present ACI framework remains an open question for future investigation. In this work, we assume that the channel is characterized sequentially using an individual probe at a time.

\subsection{Generalizability, Calibration Pipelines, and Dynamic Deployment}

We conclude by outlining potential extensions and several considerations for practical deployment. First, the combined use of segmentation and distortion redistribution for characterization and compensation injection is potentially generalizable to other unitary, square-integrable transforms (e.g., fractional Fourier, Mellin, Hilbert, or Hankel). Furthermore, by matching the unitary kernel to the underlying channel geometry, the ACI framework could be adapted for log-polar imaging, edge enhancement, and the generation of non-diffracting beams or cylindrical wave systems \cite{nguyen2025demand,wu2024method,rasouli2025fractional,traver2010review}. The selection of established kernels or the targeted development of customized ones within this framework may present significant advantages and warrants further investigation. 

Second, the segmented characterization protocol serves as a one-time, pre-deployment calibration to identify the optimal functional mapping for $\mathcal{P}_j(\mathbf{k})$ in Eq.~\ref{eq:CISMapping}. Depending on the statistical complexity of the distortion, this mapping is established through one of two distinct pathways. For analytically tractable distortions, such as shot or Johnson--Nyquist noise, it is derived formally \cite{ghoshroy2018enhanced,ghoshroy2018hyperbolic,adams2018plasmonic}. For stochastically complex regimes, such as strongly turbulent media, it is obtained via iterative calibration with known reference targets \cite{ghoshroy2024enhancing,moazzam2025remote}. Both pathways yield a deterministic CIS estimate for each probe. The estimated mapping in Eq.~\ref{eq:CISMapping} is then rigorously validated using fidelity metrics \cite{ghoshroy2018hyperbolic,ghoshroy2024enhancing}, during which both the initial trial segmentation and the functional mapping are optimized.

Third, the validation stage defines the operational boundary within which the estimated CIS maintains its fidelity. This reliability range fundamentally exploits the energy redistribution properties of square-integrable unitary transforms, which enable distortion compensation while preventing the CIS itself from introducing additional aberrations. When the propagation channel deviates from this calibrated regime, the system operates outside its optimal domain, and performance degradation is expected, as the injected CIS may introduce residual distortions rather than suppress them \cite{adams2018plasmonic,ghoshroy2024enhancing}.

Finally, in scenarios where the propagation medium evolves spatiotemporally, a dynamic operational update to the CIS is required. During deployment, this translates to propagating a set of $N+1$ probes and performing a coherent measurement of the resulting fields. The reconstructed probes and their reference counterparts are then directly substituted into the previously determined functional mappings to obtain updated CIS estimates. Consequently, no iterative real-time optimization or adaptive feedback loop is required during deployment \cite{ghoshroy2024enhancing}. This could enable easier integration with highly dynamic complex channels, provided the CIS can be estimated within the speckle decorrelation time, and the channel does not significantly deviate from the calibrated domain.  

\subsection{Extending Correlation-Injecting Sources to Unknown Optical Fields}

We now consider the more challenging scenario of propagating optical fields with unknown or complex spatial mode profiles through the random media with minimal distortion. We first examine how the set of $N+1$ CISs in Eq.~\ref{eq:RhoSet} compensates for distortion to fields whose modal structure is not known a priori. Consider an optical mode $o(\mathbf{r})$ with unknown spatial geometry and Fourier transform pair $O(\mathbf{k}) = \mathcal{F}\{o(\mathbf{r})\}$. We assume that the PSD of $O(\mathbf{k})$ is confined within the operational bandwidth $B$ of the system. While $\vert{}O(\mathbf{k})\vert{}^2$ need not span the entire bandwidth $B$, its energy must remain well-contained. Significant spectral energy outside $B$ forces the system beyond its characterized regime, as no CISs are defined for spatial frequencies $k > B/2$. 

In practical deployment, diffraction, severe beam wander, pointing jitter, or tilt instabilities may violate this assumption by introducing asymmetric aperture-related truncation effects, particularly as the operational bandwidth approaches the physical aperture limits in the Fraunhofer plane \cite{ghoshroy2024enhancing}. For this discussion, however, we assume that the spectral containment condition remains satisfied. When transmitting targets with arbitrary spectral distributions, retaining the same spectral segmentation employed during CIS characterization may be advantageous, though it is not strictly required \cite{ghoshroy2018enhanced,ghoshroy2018hyperbolic}. For clarity and without loss of generality, we assume identical spectral segmentation for the remainder of this section.
  
An unknown target field is first decomposed into a set of $N+1$ partial targets.  The $j^\mathrm{th}$ partial target is defined as $o_j(\mathbf{r}) = o(\mathbf{r}) \ast v_j(\mathbf{r})$ and therefore, the original target is expressed as the superposition
\begin{equation}
o(\mathbf{r}) = \mathlarger{\mathlarger{\sum}}_{j=0}^{N} o_j(\mathbf{r}) = \mathlarger{\mathlarger{\sum}}_{j=0}^{N} o(\mathbf{r}) \ast v_j(\mathbf{r}).
\label{eq:TotalTarget}
\end{equation}
Because characterization yields a set of $N+1$ CISs over the operational bandwidth, any target field confined within this spectral support admits compensation through the same reciprocal-space framework. Similar to Eq.~\ref{eq:InjectedProbe}, the estimated CIS corresponding to the $j^{\mathrm{th}}$ spectral segment is injected into each partial target via convolution, yielding the correlation-injected partial target $o_{j,A}(\mathbf{r})$, expressed as
\begin{equation}
o_{j,A}(\mathbf{r}) = o_j(\mathbf{r}) \ast \rho_j(\mathbf{r}) = o(\mathbf{r}) \ast v_j(\mathbf{r}) \ast \rho_j(\mathbf{r}).
\label{eq:InjectedPartialTarget}
\end{equation}
Consequently, the overall correlation-injected target is defined based on Eq.~\ref{eq:TotalTarget} as the superposition of $N+1$ partial targets as
\begin{equation}
o_A(\mathbf{r}) = \mathlarger{\mathlarger{\sum}}_{j=0}^{N} o_{j,A}(\mathbf{r})
= \mathlarger{\mathlarger{\sum}}_{j=0}^{N} o(\mathbf{r}) \ast v_j(\mathbf{r}) \ast \rho_j(\mathbf{r}).
\label{eq:InjectedTarget}
\end{equation}
This is illustrated schematically in Fig.~\ref{fig:ACIFullSchematic_partC}. The modification introduced to the original target is termed the auxiliary source $a(\mathbf{r})$, allowing the injected field in Eq.~\ref{eq:InjectedTarget} to be expressed as $o_A(\mathbf{r}) = o(\mathbf{r}) + a(\mathbf{r})$, where $a(\mathbf{r})$ is correlated with both the original target $o(\mathbf{r})$ and the ensemble of $N+1$ CISs. To ensure that each spectral segment remains decoupled from the stochastic energy of its neighbors, the partial targets $o_{j,A}(\mathbf{r})$ are propagated through the distorting medium sequentially. During propagation, the auxiliary source compensates for accumulated distortion and ideally vanishes at the detection plane, yielding a higher-fidelity realization of the target field as shown in Fig.~\ref{fig:ACIFullSchematic_partA}.

After propagating through the distorting medium, each correlation-injected partial target in Eq.~\ref{eq:InjectedTarget} is reconstructed to eliminate diffraction-induced phase advance, yielding the $j^\mathrm{th}$ reconstructed partial target $\tilde{o}_{j,A}(\mathbf{r})$. The resulting ensemble of $N+1$ reconstructed partial targets is recombined to yield the total propagated field $\tilde{o}_A(\mathbf{r})$. Because the original target was spectrally segmented via Eq.~\ref{eq:TotalTarget}, this recombination performs the inverse desegmentation process. Each component in the ensemble is inverse-weighted via convolution with $v_j(\mathbf{r})$, and the spectrally reweighted fields are coherently superimposed to restore the ground truth. Consequently, the reconstructed target is expressed as \cite{ghoshroy2024enhancing}
\begin{equation}
\tilde{o}_A(\mathbf{r}) = \mathlarger{\mathlarger{\sum}}_{j=0}^{N} w_{j}(\mathbf{r})\ast \bigg \{ \tilde{o}_{j,A}(\mathbf{r}) \ast \frac{1}{1+v_j(\mathbf{r})} \bigg \},
\label{eq:TargetReconstructed}
\end{equation}
where $w_j(\mathbf{r})$ denotes a window function tailored to the functional form of $v_j(\mathbf{r})$, introduced to account for non-ideal distortion compensation by the injected CIS \cite{ghoshroy2024enhancing,ghoshroy2018enhanced}. Since $v_j(\mathbf{r})$ is a band-limited function in reciprocal space, any residual energy in $\tilde{O}_{j,A}(\mathbf{k})$ outside the $j^\mathrm{th}$ spectral segment is assumed to be residual distortion and is removed using $w_j(\mathbf{r})$. Note that the term $1+v_j(\mathbf{r})$ in Eq. \ref{eq:TargetReconstructed} prevents unintentional magnification of distortion outside the $j^\mathrm{th}$ spectral segment before windowing. Therefore, Eq. \ref{eq:TargetReconstructed} is designed to mitigate the impact of imperfect compensation in $\tilde{o}_{j,A}(\mathbf{r})$. Any residual distortion is confined to the corresponding $j^\mathrm{th}$ spectral segment. 

It is instructive to consider scenarios where distortion compensation may be compromised under practical deployment conditions. Note that the primary objective is to ensure that $\mathbb{E}\!\left[\,|\tilde{o}_A(\mathbf{r})|^2\,\right] \approx |o(\mathbf{r})|^2$. Thus, the mode profile of the reconstructed target converges toward that of the ground truth despite imperfect compensation. 

First, the ACI framework imposes no prior constraints on the physical extent of the target field $o(\mathbf{r})$. Each CIS $\rho_j(\mathbf{r})$ encodes a spatiotemporal snapshot of the conjugate channel-induced aberration within the diffraction cone of the corresponding $j^{\text{th}}$ probe. However, the target's unknown spatial extent and power spectral density generally produce a physical-space mode profile $\vert{}o_{j,A}(\mathbf{r})\vert{}^2$ in Eq.~\ref{eq:InjectedPartialTarget} that differs from the probe profile $\vert{}v_j(\mathbf{r})\vert{}^2$. As the disparity between probe and target mode sizes grows, their effective propagation paths and phase evolution diverge from calibration conditions, degrading CIS optimality and lowering achievable compensation. Such path-dependent disparities are particularly pronounced in spatiotemporally varying channels, such as turbulent or turbid media. This issue is further exacerbated when the underlying channel exhibits anisoplanatism  (i.e., light fields arriving from different angular directions experience different optical distortions, even when they pass through the same propagation medium). However, as demonstrated in \cite{ghoshroy2024enhancing}, compensation fidelity remains robust for targets with spatial extents up to roughly twice the isoplanatic patch size  (i.e., angular range over which the wavefront aberrations are approximately identical for light arriving from different directions), even when the target's intrinsic spatial resolution exceeds the system's nominal resolution limit by more than an order of magnitude. Accordingly, anisoplanatism limits the field of view over which the calibrated CIS ensemble can compensate targets whose spectra are confined within the operational bandwidth $B$ while maintaining the required reconstruction fidelity.

The above challenge can be mitigated by explicitly accounting for anisoplanatic behavior during the characterization stage when formulating the optimal functional mapping for $\rho_j(\mathbf{r})$ in Eq.~\ref{eq:RhoSet}. Several strategies were proposed in \cite{ghoshroy2024enhancing}. For example, rather than relying on a single probe per spectral segment, controllable degrees of freedom, such as beam tilt or diameter, can be incorporated during characterization to accommodate anisoplanatism. Consequently, probe design emerges as an intrinsic component of the ACI framework. Quantitative fidelity metrics, including NCC, SSIM, or PSNR, can then be used to calibrate the optimal imaging field of view where distortion mitigation remains acceptable. 

In brief, ACI introduces a fundamentally different distortion compensation strategy from conventional approaches (e.g., adaptive optics, wavefront shaping, and optical phase conjugation), by leveraging reciprocal-space segmentation, spectrally localized distortion characterization, and convolutional compensation injection to achieve distinctive advantages. These features exploit the distortion redistribution properties of unitary integral transforms to enable reliable distortion mitigation in high-resolution, high-density, and strongly scattering or dynamic environments where conventional correction methods reach their limits. It is worth noting that the ACI framework remains compatible with hybridization with other techniques. For example, the reconstruction method presented with Eq.~\ref{eq:TargetReconstructed} could be supplemented with more sophisticated deep-learning based approaches \cite{moazzam2026active}. Similarly, distortion estimation via guide stars or pilot beams is a rapidly evolving field of research, particularly in applications involving long-distance atmospheric propagation and deep-tissue imaging through scattering, turbulent live biological media. Combining traditional physical-space correction with ACI's reciprocal-space characterization and correlation injection could propel hybrid techniques into challenging regimes unattainable by either method alone.

\subsection{Interferometric Estimation of the Complex Probe Field for CIS Generation}

We now direct our attention to the coherent measurement of the complex optical field, which is necessary for evaluating the CIS. We note that both the auxiliary source and the CIS are assumed to remain spatially and temporally coherent with respect to the ground truth field. While ACI has previously been investigated under shot-noise-limited near- and far-field incoherent imaging conditions \cite{adams2018plasmonic,adams2017plasmonic}, extensions to more complex propagation environments, such as turbulent atmospheric channels, remain an open area of investigation. Nevertheless, the potential influence of spatially and temporally quasi-coherent sources on ACI performance has been examined in \cite{ghoshroy2024enhancing}, providing preliminary insight into the robustness of the framework under partial coherence conditions.

\begin{figure}[htbp]
\includegraphics[width=0.95\linewidth]{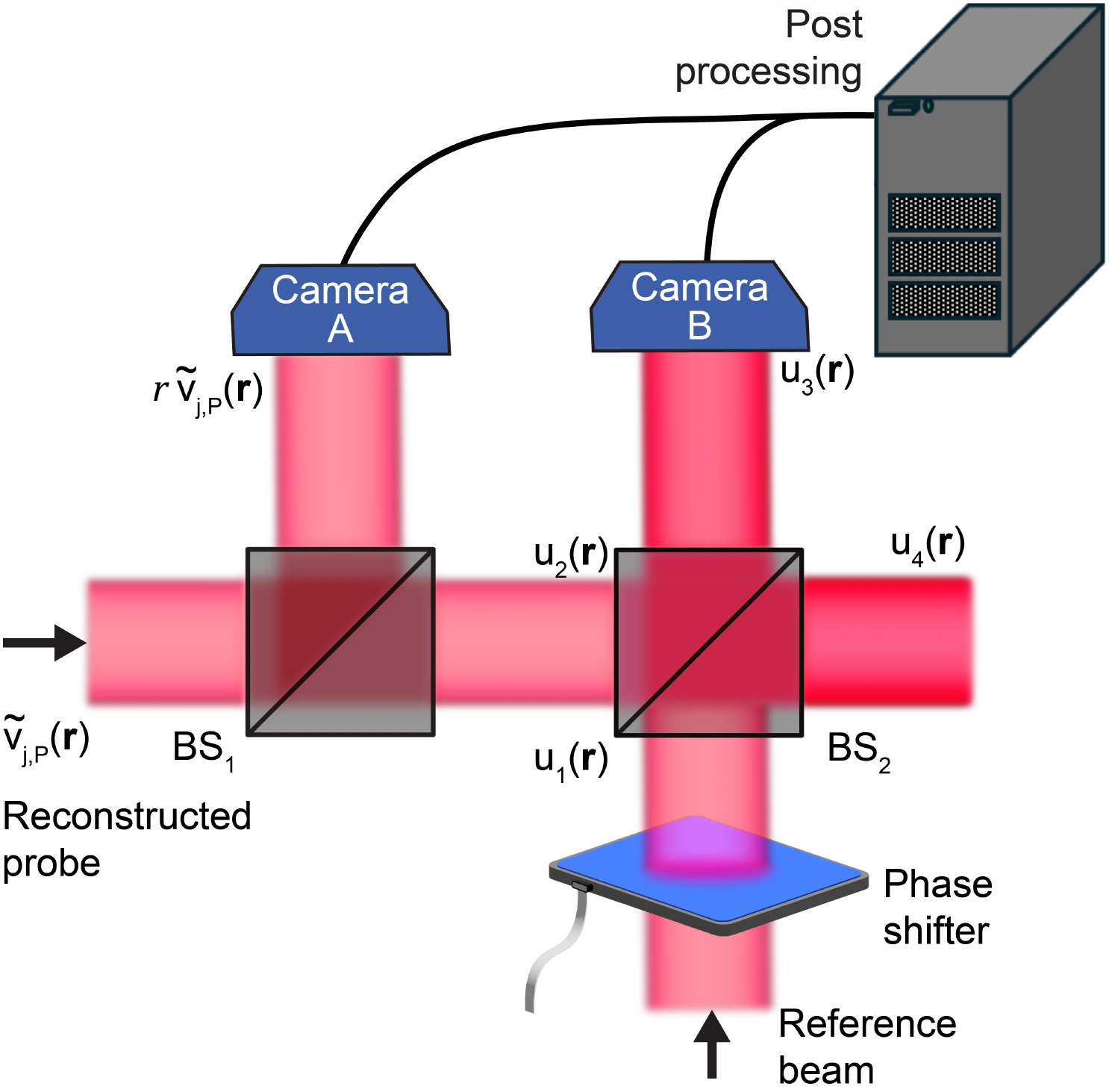}
\caption{Receiver-side phase-shifting interferometric scheme for recovering the complex field of the reconstructed ACI probe $\tilde{v}_{j,P}(\mathbf{r})$. Beam splitter $\mathrm{BS}_1$ directs a portion, $|r|^2=1/2$, of the probe to the first camera for measurement of its spatial intensity distribution, while the remaining field $u_2(\mathbf{r})$ is transmitted to the 50:50 beam splitter $\mathrm{BS}_2$. There, it interferes with a known reference field $u_1(\mathbf{r})$ whose phase is controlled by a phase shifter. The interferometric intensity output corresponding to $u_3(\mathbf{r})$ is recorded by the second camera for reference phases $\theta=0$ and $\theta=\pi/2$, while the complementary output $u_4(\mathbf{r})$ exits through the unused port. Post-processing combines the direct-intensity and phase-shifted interferometric measurements to reconstruct the amplitude and spatial phase of the probe field.}
\label{fig:DemoACIPhaseExrtraction}
\end{figure}

To illustrate how the required complex-field information may be obtained in a practical setting, we consider a receiver-side phase-shifting interferometric arrangement that separately recovers the amplitude and phase of the received field. The proposed configuration is illustrated in Fig.~\ref{fig:DemoACIPhaseExrtraction}. The reconstructed probe beam, $\tilde{v}_{j,P}(\mathbf{r})$ is incident on an ideal 50:50 beamsplitter $\mathrm{BS}_1$. The transmitted field is given by $u_2(\mathbf{r})=\tilde{v}_{j,P}(\mathbf{r})/\sqrt{2}$, while the reflected output is directed to camera A to measure the spatial intensity distribution, $|\tilde{v}_{j,P}(\mathbf{r})|^2/2$, from which the probe amplitude can be recovered. The field $u_2(\mathbf{r})$ is subsequently transmitted to a second 50:50 beamsplitter $\mathrm{BS}_2$, where it interferes with a known reference field $u_1(\mathbf{r})$ whose phase is controlled by a phase shifter. The resulting output intensity distribution $|u_3(\mathbf{r})|^2$ is recorded by a second camera B, while $u_4(\mathbf{r})$ exits through the unused port. The measurements from both cameras are subsequently post-processed to reconstruct the spatially varying phase of the probe field.

The phase information of the reconstructed probe beam can be extracted from the input-output relations of the beam splitters. We assume $\mathrm{BS}_1$ and $\mathrm{BS}_2$ are two-input, lossless, 50:50 beam splitters. The complex output fields $u_3(\mathbf{r}) \in \mathbb{C}$ and $u_4(\mathbf{r})\in \mathbb{C}$ in Fig.~\ref{fig:DemoACIPhaseExrtraction} are related to the complex input fields $u_1(\mathbf{r}) \in \mathbb{C}$ and $u_2(\mathbf{r}) \in \mathbb{C}$ as
\begin{equation}
    \begin{bmatrix}
        u_3(\mathbf{r}) \\
        u_4(\mathbf{r})
    \end{bmatrix}
    = \frac{1}{\sqrt{2}}
    \begin{bmatrix}
        1 & e^{i\frac{\pi}{2}} \\
        e^{i\frac{\pi}{2}} & 1
    \end{bmatrix}
    \begin{bmatrix}
        u_1(\mathbf{r}) \\
        u_2(\mathbf{r})
    \end{bmatrix},
\label{eq:BeamSplitterEq}    
\end{equation}
which upon expansion yields 
\begin{subequations} 
\begin{align}
u_3(\mathbf{r}) &= \frac{1}{\sqrt{2}}\left( u_1(\mathbf{r})+u_2(\mathbf{r})e^{i\frac{\pi}{2}}\right), \\ 
u_4(\mathbf{r}) &= \frac{1}{\sqrt{2}}\left( u_1(\mathbf{r})e^{i\frac{\pi}{2}} +u_2(\mathbf{r}) \right).
\end{align}
\label{eq:BSIO_Eqns}
\end{subequations}
The reference beam incident on $\mathrm{BS}_2$ is taken to be a plane wave, $u_1(\mathbf{r})=|p|e^{i\theta}/\sqrt{2}$, where $|p|/\sqrt{2}$ and $\theta$ denote its amplitude (for notational convenience) and phase, respectively. Similarly, the second input to $\mathrm{BS}_2$ is the reconstructed probe field $u_2(\mathbf{r})=|\tilde{v}_{j,P}(\mathbf{r})|e^{i\phi_{j,P}(\mathbf{r})}/\sqrt{2}$, where $\phi_{j,P}(\mathbf{r})$ is the unknown phase of the $j^{\mathrm{th}}$ reconstructed probe. Substituting these expressions into the relation for $u_3(\mathbf{r})$ in Eq.~\ref{eq:BSIO_Eqns}(a) and using $e^{i\pi/2}=i$ gives
\begin{align} 
u_3(\mathbf{r}) &= \frac{1}{2}\left( |p|e^{i\theta} +|\tilde{v}_{j,P}(\mathbf{r})|e^{i\phi_{j,P}(\mathbf{r})}e^{i\pi/2} \right) \notag \\ 
&= \frac{1}{2}\left( |p|e^{i\theta} +i|\tilde{v}_{j,P}(\mathbf{r})|e^{i\phi_{j,P}(\mathbf{r})} \right). 
\label{eq:BSOP_u3}
\end{align}
By adjusting the phase imparted by the phase shifter, the reference-beam phase $\theta$ can be sequentially set to $0$ and $\pi/2$. The corresponding spatial interference patterns recorded at the camera are denoted by $\left.|u_3(\mathbf{r})|^2\right|_{\theta=0}$  and $\left.|u_3(\mathbf{r})|^2\right|_{\theta=\pi/2}$, respectively. Evaluating Eq.~\ref{eq:BSOP_u3} at these two reference phases and applying Euler's formula, the corresponding interference patterns are expressed as
\begin{align}
\left.|u_3(\mathbf{r})|^2\right|_{\theta=0} = \frac{1}{4}\big(|p|^2 + & |\tilde{v}_{j,P}(\mathbf{r})|^2  \notag \\ & - 2|p||\tilde{v}_{j,P}(\mathbf{r})|\sin \phi_{j,P}(\mathbf{r})\big)
\label{eq:BSOP_u3_theta0}
\end{align}
and 
\begin{align}
\left.|u_3(\mathbf{r})|^2\right|_{\theta=\pi/2} = \frac{1}{4} \big( |p|^2 + & |\tilde{v}_{j,P}(\mathbf{r})|^2 \notag \\ & + 2|p||\tilde{v}_{j,P}(\mathbf{r})|\cos \phi_{j,P}(\mathbf{r}) \big)
\label{eq:BSOP_u3_thetaPi2}, 
\end{align}
respectively. Note that Equations~\ref{eq:BSOP_u3_theta0} and \ref{eq:BSOP_u3_thetaPi2} can be rearranged to obtain the sine and cosine quadratures of the unknown spatial phase $\phi_{j,P}(\mathbf{r})$, respectively. Neither quadrature independently determines the phase over the full $2\pi$ interval because a given sine or cosine value may correspond to multiple phase angles. Together, however, the two quadratures contain the information required to resolve the correct phase quadrant. Isolating the sine and cosine terms in Equations~\ref{eq:BSOP_u3_theta0} and \ref{eq:BSOP_u3_thetaPi2} and taking their ratio gives 
\begin{align}
\tan\phi_{j,P}(\mathbf{r}) &= \frac{
|p|^2+|\tilde{v}_{j,P}(\mathbf{r})|^2
-4\left.|u_3(\mathbf{r})|^2\right|_{\theta=0}
}{
4\left.|u_3(\mathbf{r})|^2\right|_{\theta=\pi/2}
-|p|^2-|\tilde{v}_{j,P}(\mathbf{r})|^2
}
\notag \\
&=
\frac{
i_{\mathrm{dc}}(\mathbf{r})
-\left.|u_3(\mathbf{r})|^2\right|_{\theta=0}
}{
\left.|u_3(\mathbf{r})|^2\right|_{\theta=\pi/2}
-i_{\mathrm{dc}}(\mathbf{r})
}, 
\label{eq:PhaseTangent}
\end{align}
where $i_{\mathrm{dc}}(\mathbf{r})=\left(|p|^2+|\tilde{v}_{j,P}(\mathbf{r})|^2\right)/4$ denotes the phase-independent contribution to each detected intensity. Eq.~\ref{eq:PhaseTangent} may therefore be used to numerically estimate the spatial phase distribution $\phi_{j,P}(\mathbf{r})$ of the reconstructed probe beam $\tilde{v}_{j,P}$ purely from measured intensity data. It is instructive to note that the accuracy of the reconstructed phase is governed primarily by the signal-to-noise ratio of the phase-dependent interference modulation. The relative probe and reference amplitudes should therefore be selected to provide high fringe visibility while avoiding detector saturation. Direct evaluation of Eq.~\ref{eq:PhaseTangent} can become numerically ill-conditioned when the cosine-based denominator approaches zero. To avoid this numerical singularity, the phase may instead be reconstructed directly from the sine and cosine quadratures without explicitly forming their ratio. Finally, physical implementation of the proposed interferometric setup requires accurate calibration of the phase shift and acquisition of both interference patterns from the same realization of the probe field.

\section{Results and Discussion}

Having outlined the operating principles of ACI, we now consider the framework from a physical deployment perspective. Applications such as free-space optical communication, biomedical imaging, and sub-diffraction imaging impose different constraints on optical access, field measurement, and propagation geometry. The generation of CISs and injection of the corresponding compensation must therefore be tailored to the system architecture and intended application.

We first consider the use of physical-space convolution for injecting CISs into an arbitrary target field, together with the recovery of phase information required for CIS estimation in coherent implementations. We then numerically evaluate the efficacy of the proposed interferometric approach for estimating the CIS. Finally, we examine two representative applications involving free-space optical communication and potential high-precision interferometric phase measurement, outlining likely pathways for experimental implementation and validation of the ACI framework.

\begin{figure*}[htbp]
\centering
\includegraphics[width=0.9\textwidth]{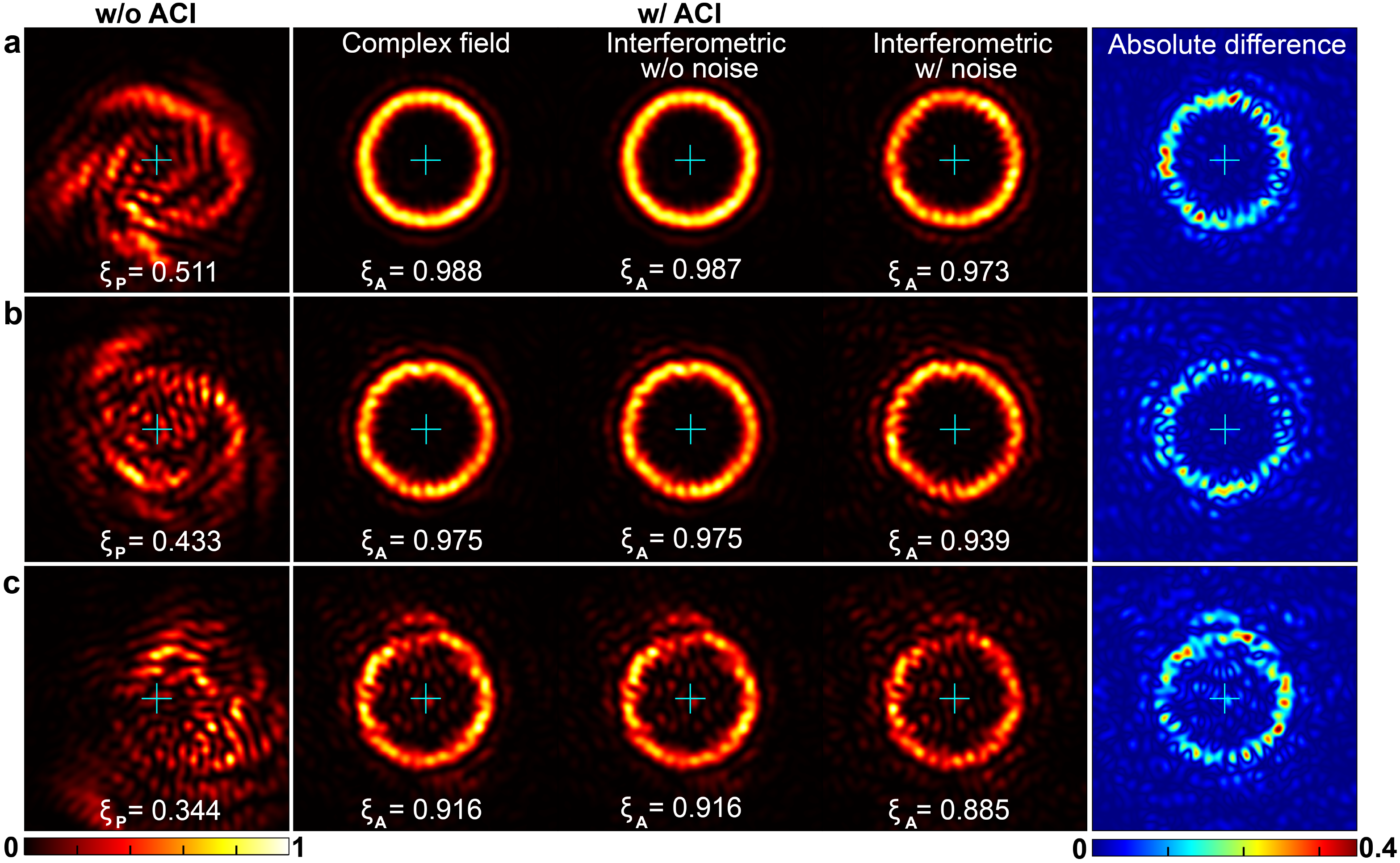}
\caption{Numerical evaluation of the proposed interferometric method for generating CISs from intensity measurements under atmospheric turbulence propagation. Rows (a)--(c) correspond to turbulence strengths $C_n^2=0.5\times10^{-14}$, $1.0\times10^{-14}$, and $1.5\times10^{-14}~\mathrm{m}^{-2/3}$, respectively. The first column shows the normalized reconstructed probe intensity without ACI. Columns 2--4 show the normalized intensities of the reconstructed correlation-injected probes using CISs estimated directly from the full complex probe field, noiseless interferometric measurements, and noisy interferometric measurements, respectively. The fifth column shows the absolute pointwise difference between the normalized intensities in columns 3 and 4, illustrating the effect of measurement noise on the reconstructed ACI-compensated OAM mode. The spatial normalized cross-correlation values $\xi_P$ and $\xi_A$, evaluated at zero lag, are indicated in the corresponding panels. Each panel covers a $3.5\times3.5~\mathrm{cm}^2$ area, with the optical axis marked by a cross. The close agreement with the complex field results shows that the interferometrically reconstructed field enables accurate CIS calculation, with only modest degradation in the presence of detector noise.}
\label{fig:CISGen_Inter}    
\end{figure*}

\subsection{Optical Access, Correlation Injection and Interferometric Phase Retrieval}

The implementation considered here assumes access to the complex probe-field information needed for CIS estimation, together with physical access to the target plane for field modulation. The latter requirement becomes restrictive when the distal plane is physically inaccessible. Representative examples include deep-tissue biological imaging, where the target may be embedded within a scattering volume \cite{gigan2022roadmap}, long-range remote sensing or non-line-of-sight (NLOS) imaging \cite{faccio2020non}, where the distal reflecting surface cannot be directly manipulated. To overcome distal access constraints, numerous feedback strategies have been developed using measurements acquired exclusively outside the sample. These approaches exploit mechanisms such as multiphoton excitation, acousto-optic/photoacoustic guide stars, and matrix decomposition algorithms \cite{bertolotti2022imaging,rumman2025optimizing,horstmeyer2015guidestar}. In reciprocal channels, the probe can be launched from the receiver side, for example, via a guide star or pilot beam offers a viable alternative, to estimate $\tilde v_{j,P}(\mathbf{r})$.

Here, we restrict our attention to scenarios in which the distal plane is directly accessible. Free-space optical (FSO) communication systems provide a particularly suitable platform because the transmitter plane can be directly accessed and manipulated. Consequently, FSO platforms offer an experimentally tractable testbed for the initial validation and benchmarking of the proposed framework under controlled conditions. Extending the framework to configurations in which the distal plane is inaccessible is left for future work. 

We next numerically evaluate the proposed interferometric method for estimating CISs from intensity measurements, assuming the setup shown in Fig.~\ref{fig:DemoACIPhaseExrtraction}. A monochromatic, coherent probe beam with wavelength $\lambda=1550~\mathrm{nm}$ is propagated horizontally over a distance $L=1~\mathrm{km}$ through a clear atmosphere free from fog, aerosols, and clouds. Wave propagation is simulated using the split-step method with ten Monte Carlo phase screens \cite{Schmidt,roggemann2018imaging,andrews2005laser}. The refractive-index structure parameter $C_n^2$, which controls the strength of turbulence induced index fluctuations, is varied over the range $0.5\times10^{-14} \leq C_n^2 \leq 1.5 \times10^{-14}~\mathrm{m}^{-2/3}$.  The propagation configuration, its numerical implementation, the functional forms of the OAM mode $v_j(\mathbf{r})$ and CIS $\rho_j(\mathbf{r})$ are identical to those employed in Ref.~\cite{ghoshroy2024enhancing}. 

The results of this analysis are presented in Fig.~\ref{fig:CISGen_Inter}, with each subfigure covering a $3.5\times3.5~\mathrm{cm}^2$ area and the optical axis is marked with a cross. The fidelity of wave propagation with and without ACI is quantified using spatial normalized cross-correlation (NCC) evaluated at zero displacement. The corresponding values are denoted by $\xi_A=|v_j(\mathbf{r})|^2\star|\tilde{v}_{j,A}(\mathbf{r})|^2$ and $\xi_P=|v_j(\mathbf{r})|^2\star|\tilde{v}_{j,P}(\mathbf{r})|^2$, respectively, where $\star$ denotes the normalized cross-correlation operator. The first column shows the normalized intensity distributions of the reconstructed probe field $\tilde{v}_{j,P}(\mathbf{r})$. Panels (a)--(c) correspond to the results for $C_n^2=0.5\times10^{-14}$, $1.0\times10^{-14}$, and $1.5\times10^{-14}~\mathrm{m}^{-2/3}$, respectively. 

As a reference, the CIS $\rho_j(\mathbf{r})$ is calculated directly from the full complex reconstructed probe fields by applying the functional mapping proposed in Ref.~\cite{ghoshroy2024enhancing}. The corresponding correlation-injected OAM mode $v_{j,A}(\mathbf{r})$ is propagated through the same turbulent channel and then reconstructed. The results are shown in the first column in Fig.~\ref{fig:CISGen_Inter} for the selected  $C_n^2$ values. Consistent with the results reported in Ref.~\cite{ghoshroy2024enhancing}, strong suppression of turbulence-induced geometric distortion is observed for all reconstructed results when compared with the results without ACI in the second column. The annular mode profile is recovered with good contrast at all three turbulence strengths, with $\xi_A$ ranging from $0.916$ to $0.988$, although residual distortion becomes more apparent at the highest $C_n^2$.

The simulations are then repeated assuming access only to intensity measurements from the interferometric setup in Fig.~\ref{fig:DemoACIPhaseExrtraction}, with the probe amplitude obtained from $|\tilde{v}_{j,P}(\mathbf{r})|^2$ and its phase $\phi_{j,P}(\mathbf{r})$ recovered using Eq.~\ref{eq:PhaseTangent}. The CIS functional mapping remains unchanged, and identical turbulence phase-screen realizations are reused at each $C_n^2$ to isolate the effect of interferometric field reconstruction. To evaluate the influence of detector noise on the recovered phase $\phi_{j,P}(\mathbf{r})$ and, consequently, the estimated CIS, signal-dependent Poisson noise and signal-independent Johnson--Nyquist noise are introduced into the simulated intensity measurements $|\tilde{v}_{j,P}(\mathbf{r})|^2$. The correlation-injected OAM modes $v_{j,A}(\mathbf{r})$, synthesized using CISs estimated from noiseless and noisy intensity measurements, are then propagated through the same turbulent channel and subsequently reconstructed. The corresponding normalized intensity distributions are shown in the third and fourth columns of Fig.~\ref{fig:CISGen_Inter}, respectively, for the selected $C_n^2$ values. The noiseless interferometric results in the third column closely reproduce the intensity profiles and NCC values of the complex field reference in the second column despite phase wrapping. At a measurement SNR of approximately $40~\mathrm{dB}$, the results in the fourth column exhibit visible deviations from the reference while retaining the overall annular structure. The corresponding $\xi_A$ values decrease modestly to $0.973$, $0.939$, and $0.885$ for panels (a)--(c), respectively. Nevertheless, these values remain substantially greater than the corresponding passive results $\xi_P=0.511$, $0.433$, and $0.344$. These results illustrate that the interferometrically reconstructed complex field retains sufficient fidelity for accurate CIS estimation and preserves the substantial propagation enhancement provided by ACI across the selected turbulence range. The final column shows the absolute pointwise difference between the normalized intensity distributions in the third and fourth columns, quantifying the deviation of the noisy interferometric result from its noiseless counterpart. The proposed interferometric arrangement may therefore facilitate CIS estimation by recovering the complex probe field from intensity measurements without requiring computationally intensive iterative phase retrieval. We note here that physical implementation requires the interference patterns $\left.|u_3(\mathbf{r})|^2\right|_{\theta=0}$ and $\left.|u_3(\mathbf{r})|^2\right|_{\theta=\pi/2}$ used in Eq.~\ref{eq:PhaseTangent} to be acquired within a time interval short compared with the speckle decorrelation time. Practical implementation would therefore involve establishing an application-specific balance among the SLM switching rate, synchronized detector acquisition, achievable spatial resolution, and field of view \cite{luo2022high,rocha2024fast}.
 
\subsection{Representative Applications in Optical Communication and High-Precision Phase Measurement} 

Having examined the operating principles of ACI and its potential to preserve intricate spatial mode profiles during propagation through noisy and distorting channels, we now consider how this capability could be used in representative applications. Consider, for example, free-space optical (FSO) communication links, where orbital angular momentum (OAM) modes have been used alongside conventional encoding and decoding techniques to substantially increase channel capacity. While high aggregate data rates have been demonstrated in laboratory and short-range links, extending such performance to larger numbers of multiplexed modes and realistic turbulent propagation remains challenging. Turbulence-induced modal crosstalk complicates channel separation, while deformation of spatial intensity profiles presents an additional challenge for approaches that encode information directly in the mode geometry. ACI has been shown to preserve the global spatial structure of an optical field after turbulent propagation, even when residual local distortions remain. This capability could support communication schemes in which information is encoded in the collective spatial geometry of the field. A densely packed beam array provides an intuitive illustration: turbulence can distort the constituent beams and obscure their spatial arrangement, making the encoded pattern difficult to distinguish. The following numerical example illustrates how ACI can help retain the overall array geometry under the propagation conditions considered.

\begin{figure*}[htbp] 
\includegraphics[width=0.18\linewidth]{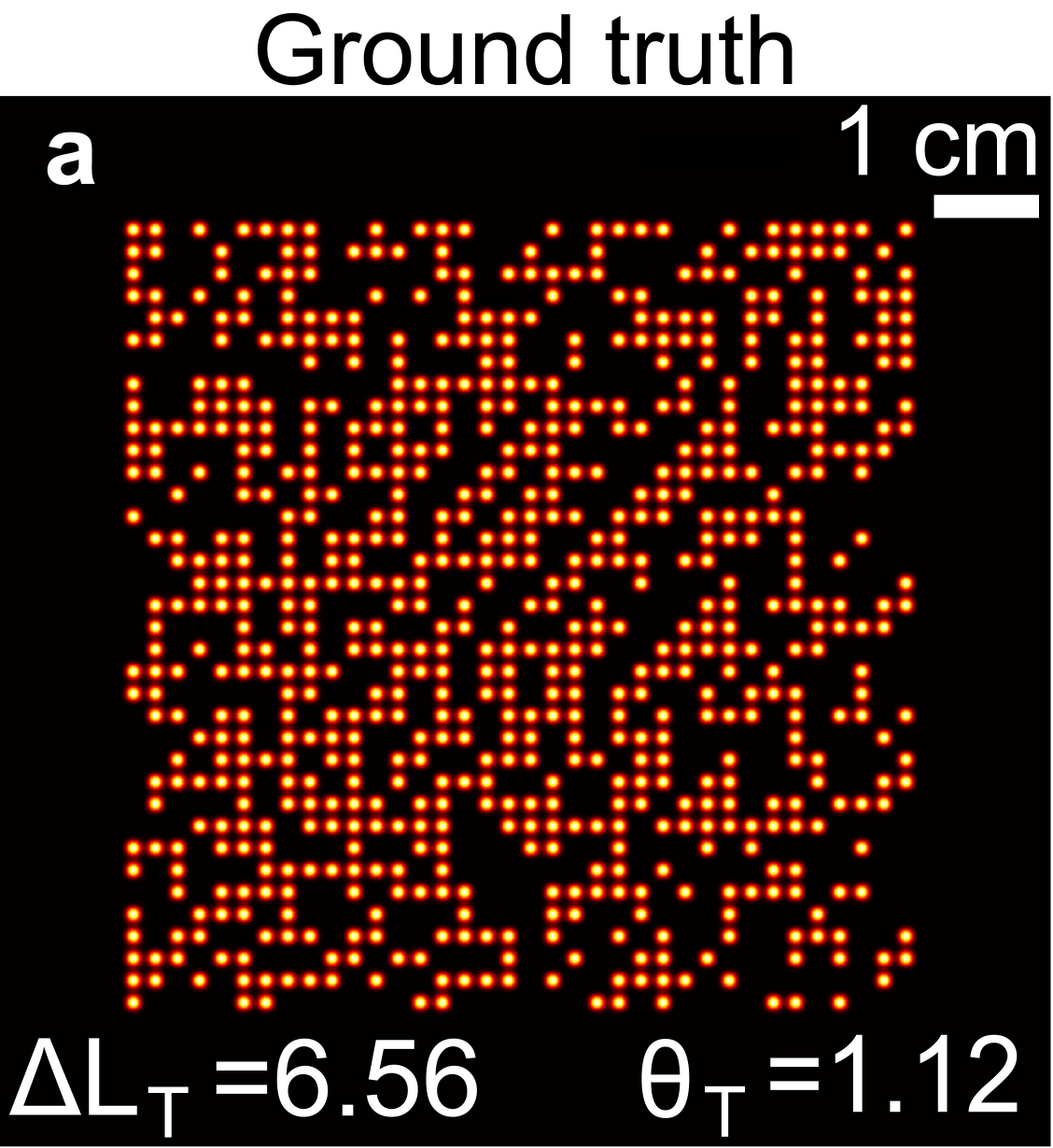}
\includegraphics[width=0.18\linewidth]{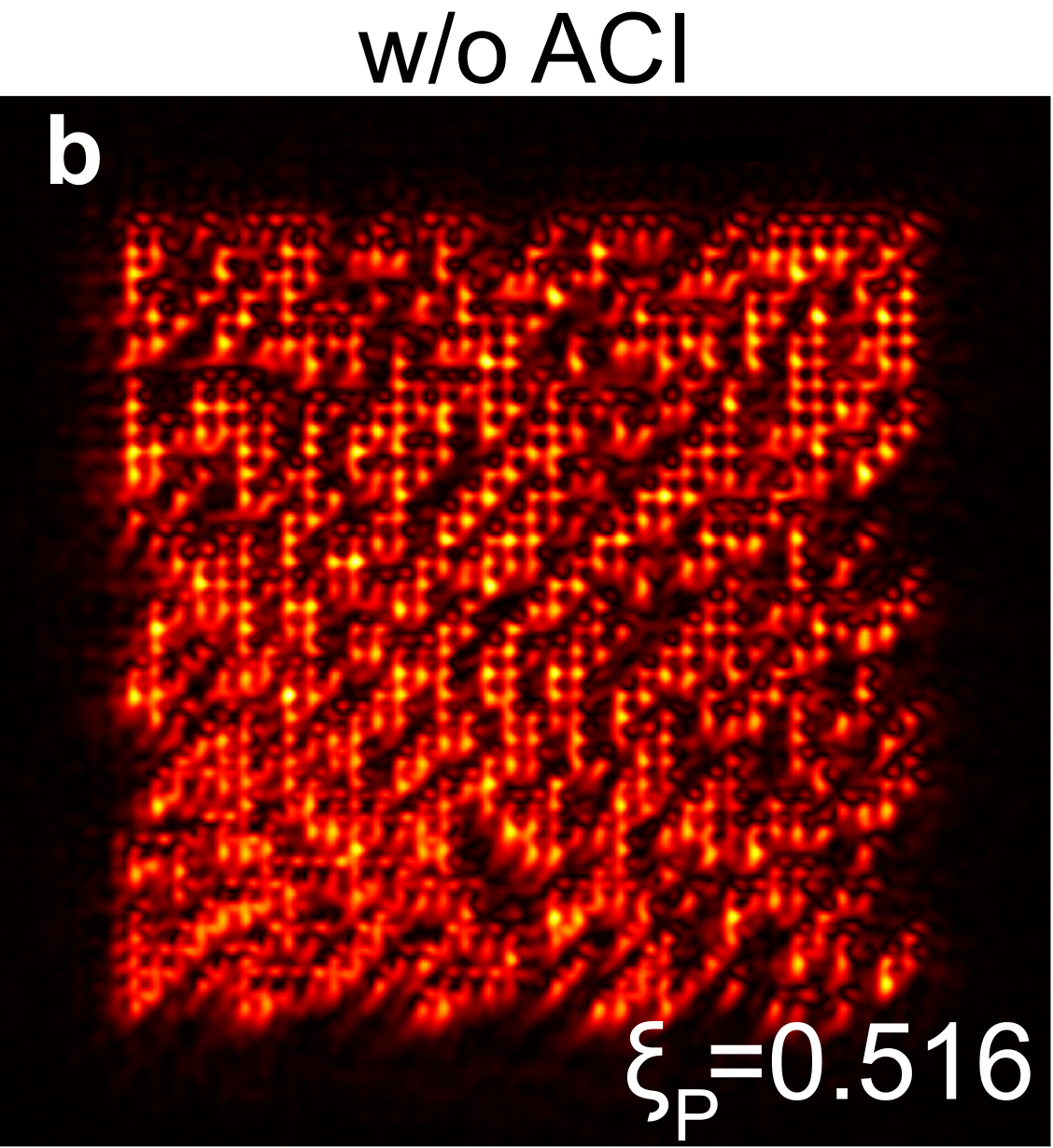}
\includegraphics[width=0.18\linewidth]{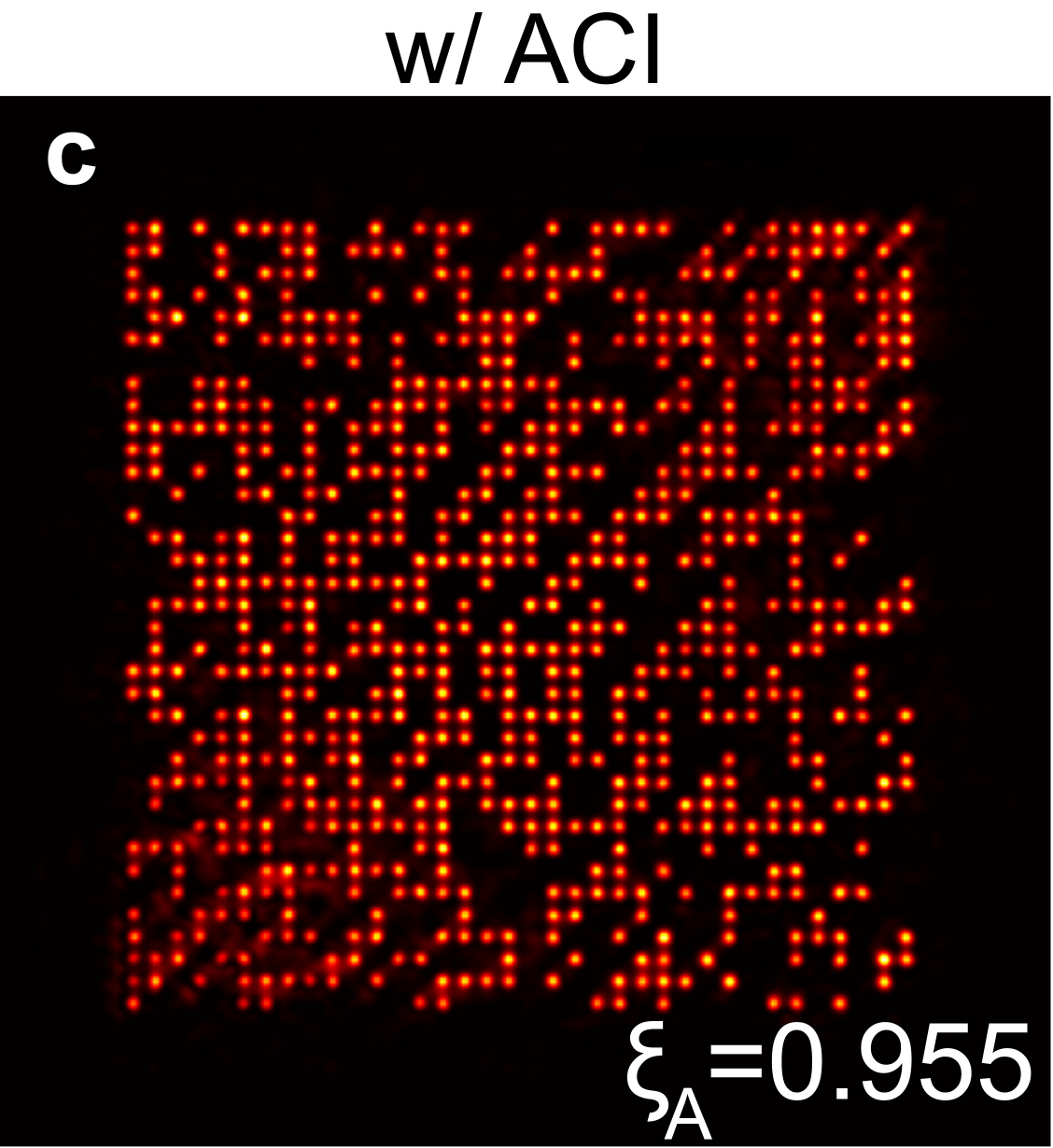}\\
\includegraphics[width=0.18\linewidth]{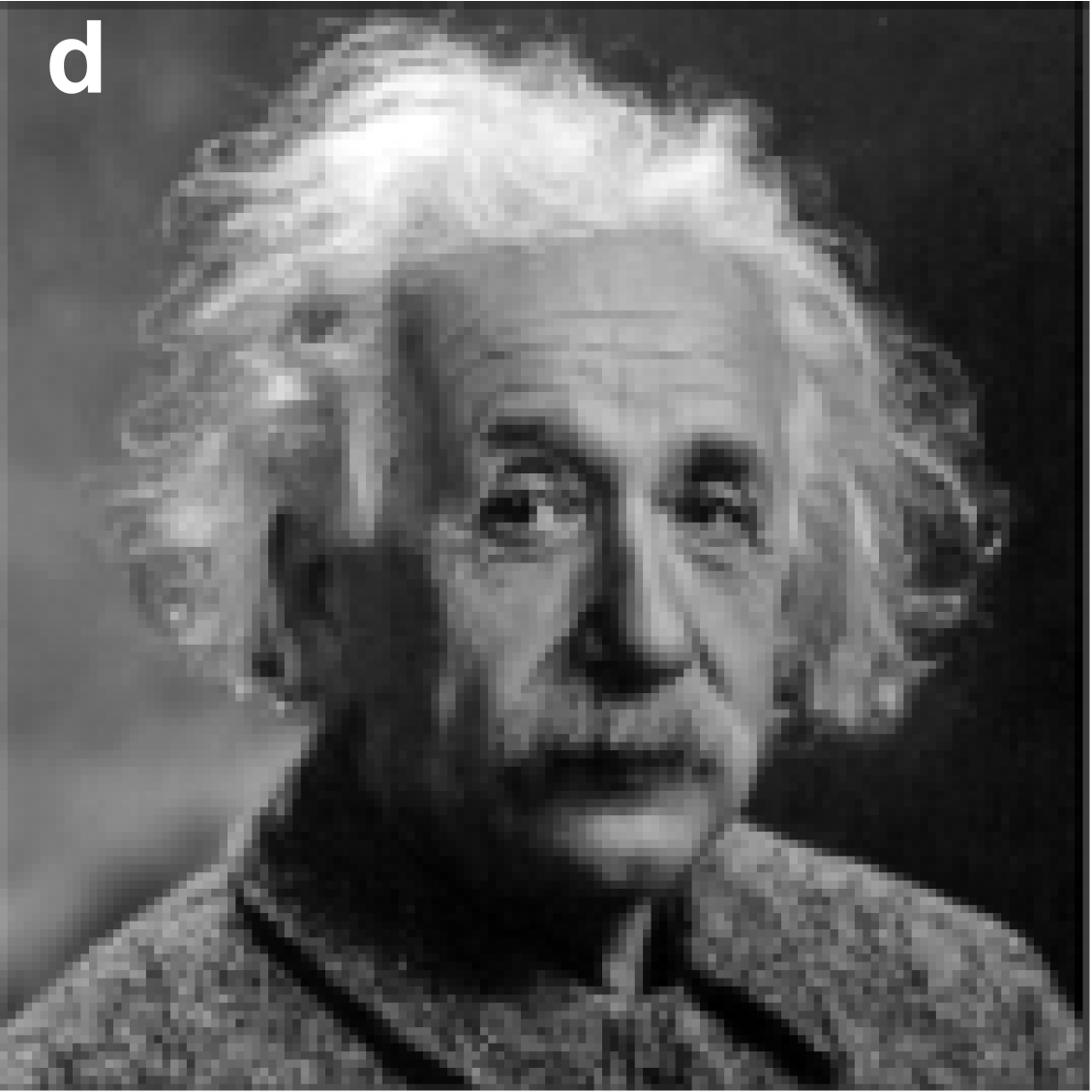}
\includegraphics[width=0.18\linewidth]{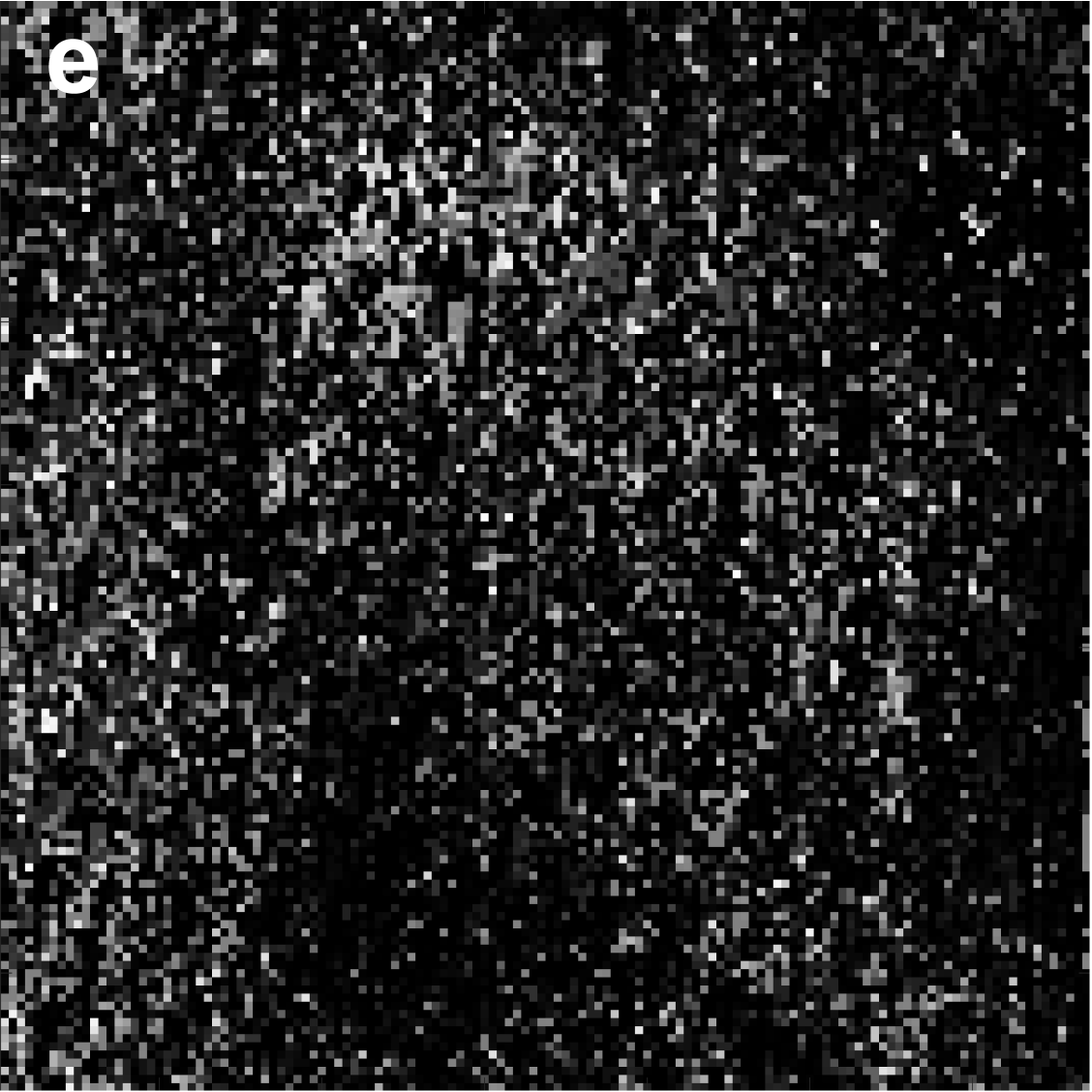}
\includegraphics[width=0.18\linewidth]{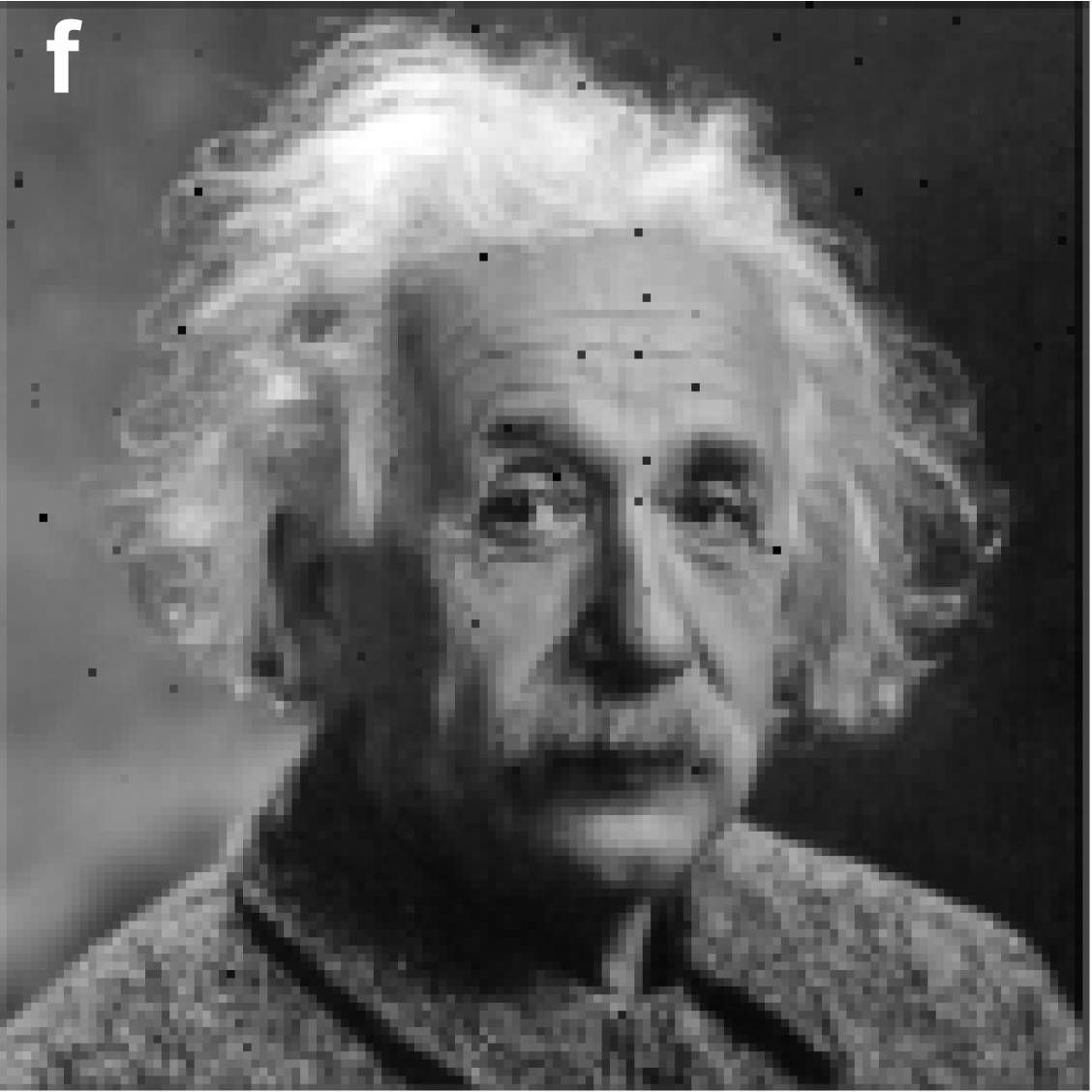}\\
\includegraphics[width=0.18\linewidth]{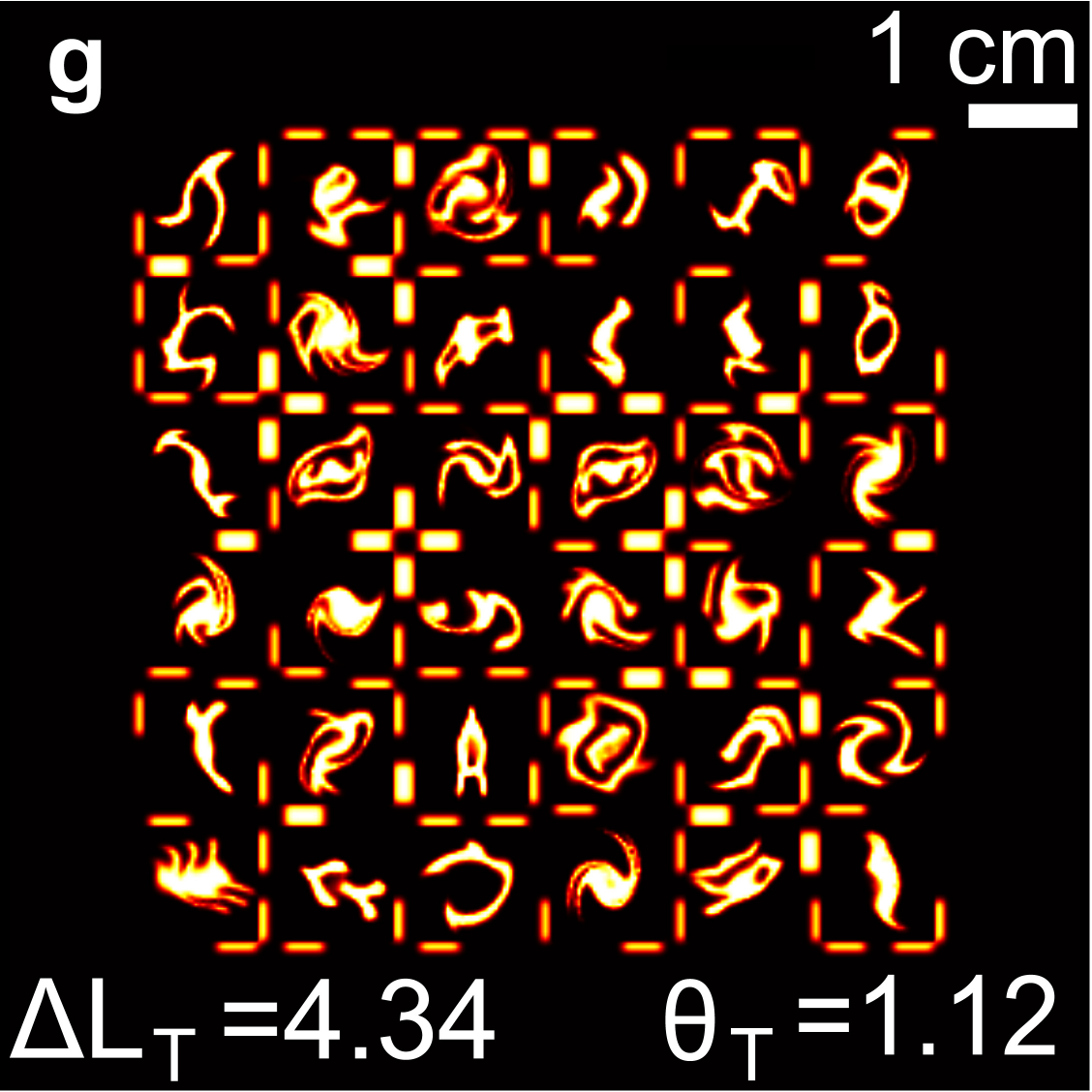}
\includegraphics[width=0.18\linewidth]{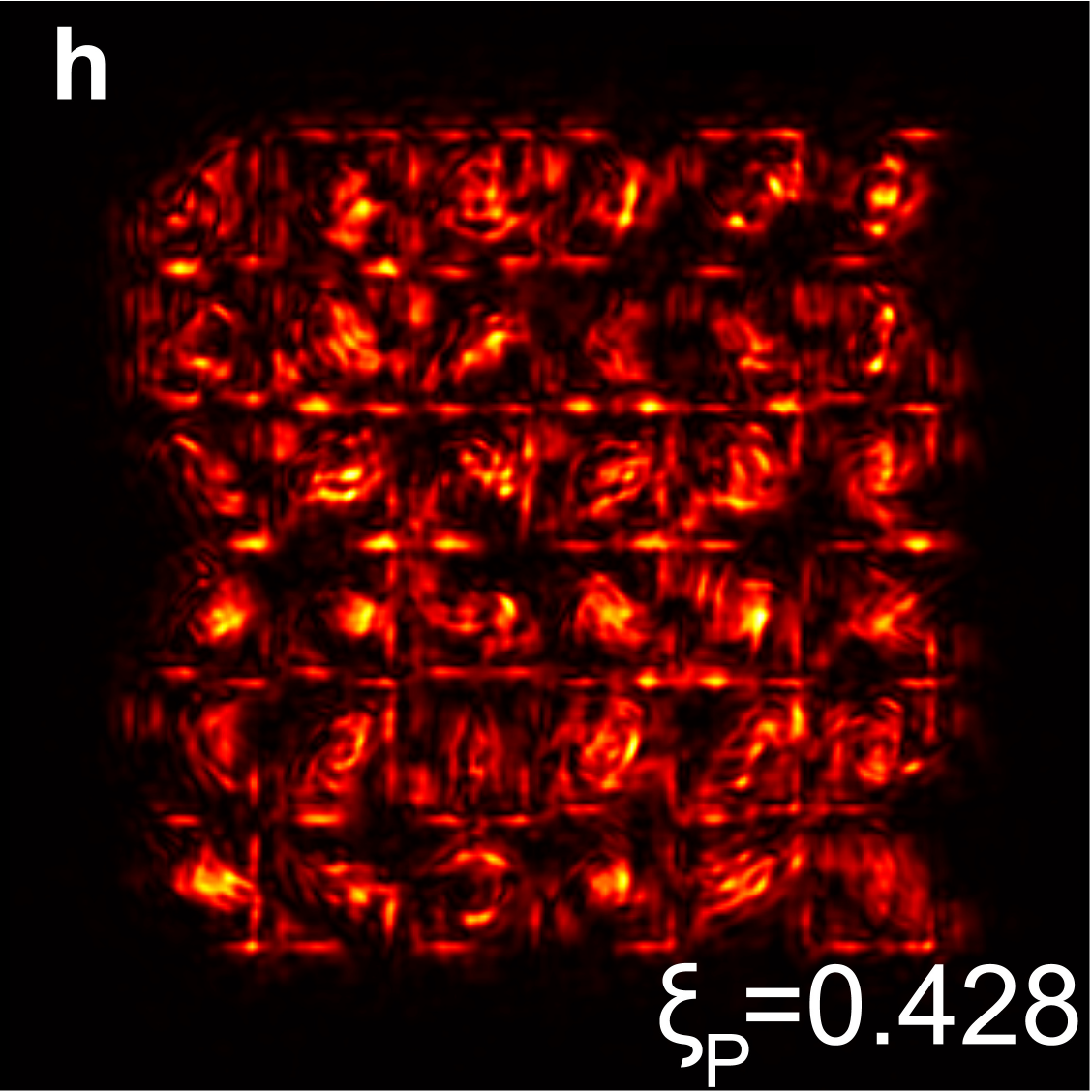}
\includegraphics[width=0.18\linewidth]{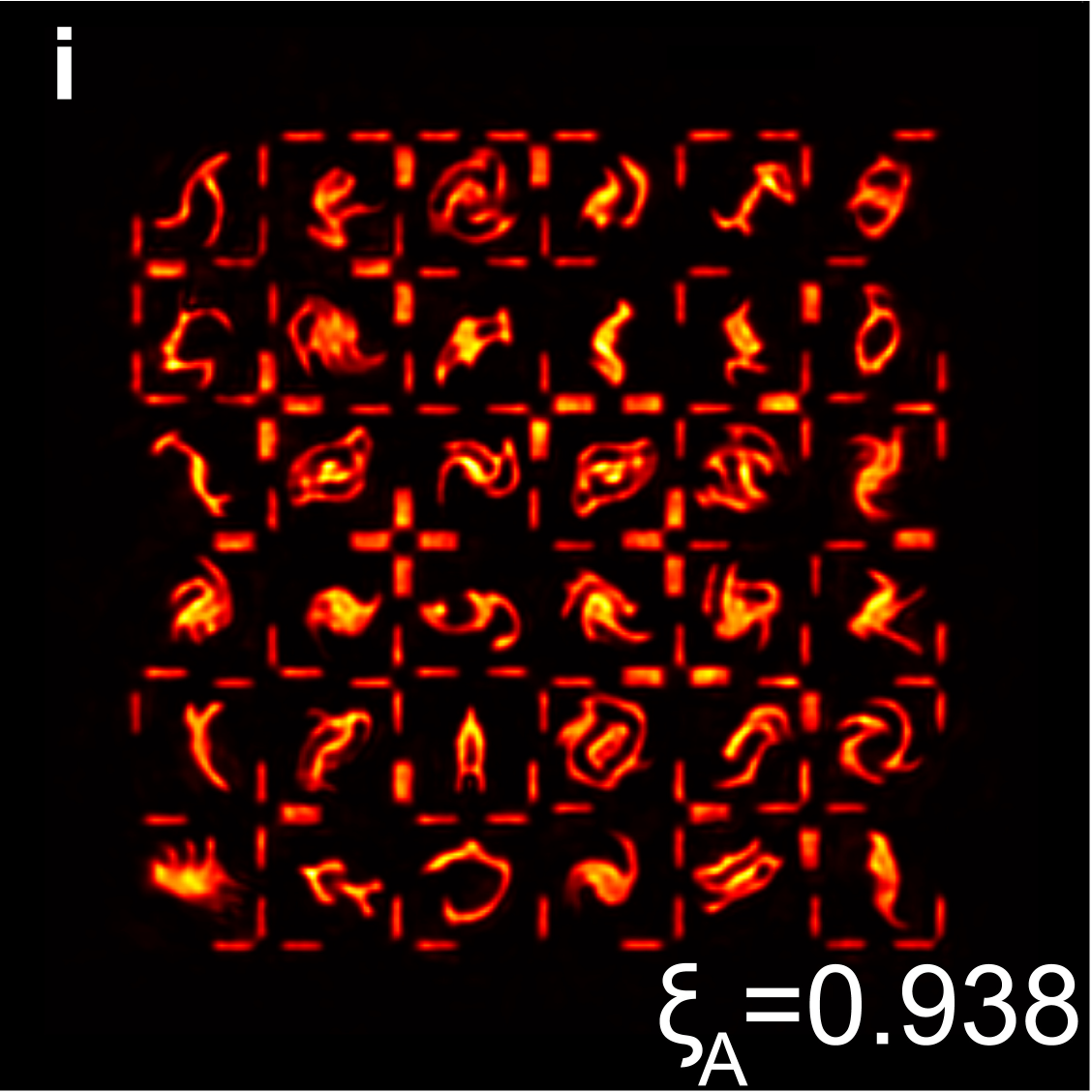}
\caption{Numerical illustration of information transmission using beam arrays and symbolic spatial encoding through atmospheric turbulence with and without ACI, for $C_n ^2 = 0.5 \times 10^{-14} \textrm{ m}^{-2/3}$. (a) An arbitrarily selected $36\times36$ beam-array frame corresponding to a 1296-bit chunk derived from the grayscale Einstein image. (b) Reconstructed normalized intensity distribution after propagation without ACI, showing substantial spatial distortion and loss of geometric fidelity. (c) Corresponding reconstruction with ACI, showing preservation of the array geometry despite residual local distortions. (d) Original $134 \times 134$ pixel grayscale reference image. (e) Image reconstructed by decoding and reassembling all 111 beam-array frames propagated without ACI, exhibiting substantial corruption. (f) Corresponding image recovered with ACI, closely reproducing the reference portrait with residual pixel-level artifacts. (g) Illustration of symbolic spatial encoding, in which information is represented by the geometry of structured optical symbols, with a nominal 1296-bit payload matching that of the beam-array frame in (a). (h) Reconstructed symbolic field after propagation without ACI, showing degradation of geometric features needed for symbol identification. (i) Corresponding reconstruction with ACI, showing preservation of the overall symbol geometry and fine features relevant to pattern recognition. The spatial normalized cross-correlation values $\xi_P$ and $\xi_A$ are indicated in the corresponding reconstructed-field panels. See text for the definitions of $\Delta L_T$ and $\theta_T$. The ground truth in (d) is adapted from a photograph of Albert Einstein by Orren Jack Turner (1947), Library of Congress, LC-USZ62-60242.}
\label{fig:DemoACIFSO}
\end{figure*}

Information transmission using a densely packed beam array is illustrated in Fig.~\ref{fig:DemoACIFSO}. A $134 \times 134$ pixel grayscale image of Albert Einstein is selected as a representative payload. Each pixel is represented by 8 bits, corresponding to 256 grayscale levels, giving a total of $134 \times134\times8=143648$ bits of uncompressed image data. The resulting bit sequence is partitioned into contiguous chunks of 1296 bits, yielding 110 complete chunks and a final chunk containing 1088 bits. The final chunk is padded with 208 trailing zeros, producing 111 chunks of equal length. Each chunk is mapped onto a $36\times36$ beam-array frame, with one bit assigned to each site of a square grid. The presence or absence of a signal represents a binary 1 or 0, respectively. Each beam corresponds to a square region of side length $6\delta_x=2.1~\mathrm{mm}$, where $\delta_x=0.35\times10^{-3}~\mathrm{m}$ is the transverse sampling interval. Fig.~\ref{fig:DemoACIFSO}(a) illustrates the mapping for an arbitrarily selected 1296-bit chunk.

To quantify channel constraints relative to target geometry, we define the dimensionless parameters, $\Delta L_T$ and $\theta_T$. The former serves as a generalized equivalent to the well-known $D/r_0$ ratio and quantifies the ratio of the finest spatial feature size of the target relative to the atmospheric resolution limit. The parameter $\theta_T = \theta_{\text{target}}/\theta_0$ quantifies the severity of channel anisoplanatism as the ratio of the target's angular footprint to the isoplanatic angle $\theta_0$. For the binary beam array in Fig.~\ref{fig:DemoACIFSO}(a), $\Delta L_T = 6.56$ indicates that the target contains features $6.56$ times finer than the atmospheric resolution limit.

Each beam-array frame is then propagated separately over a distance of $1~\mathrm{km}$ through the turbulent channel, with and without ACI. The propagation configuration is identical to that employed in Ref.~\cite{ghoshroy2024enhancing}, with the refractive-index structure parameter fixed at $C_n^2=0.5\times10^{-14}~\mathrm{m}^{-2/3}$. The normalized intensity distributions of the reconstructed beam arrays without and with ACI are shown in Figs.~\ref{fig:DemoACIFSO}(b) and (c), respectively. The corresponding bit sequences are recovered using a simple intensity-threshold decoder. The summed intensity within each beam pixel is compared with a threshold, assigning a binary 1 when the threshold is exceeded and 0 otherwise. The decoded sequences from all 111 frames are then concatenated in their original order. After removing the 208 padding bits, the remaining bits are grouped into 8-bit pixel values to reconstruct the full grayscale image. The decoded grayscale images obtained without and with ACI are shown in Figs.~\ref{fig:DemoACIFSO}(e) and (f), respectively, alongside the original reference image in Fig.~\ref{fig:DemoACIFSO}(d).  

\begin{figure*}[htbp]
\centering
\includegraphics[width=0.96\textwidth]{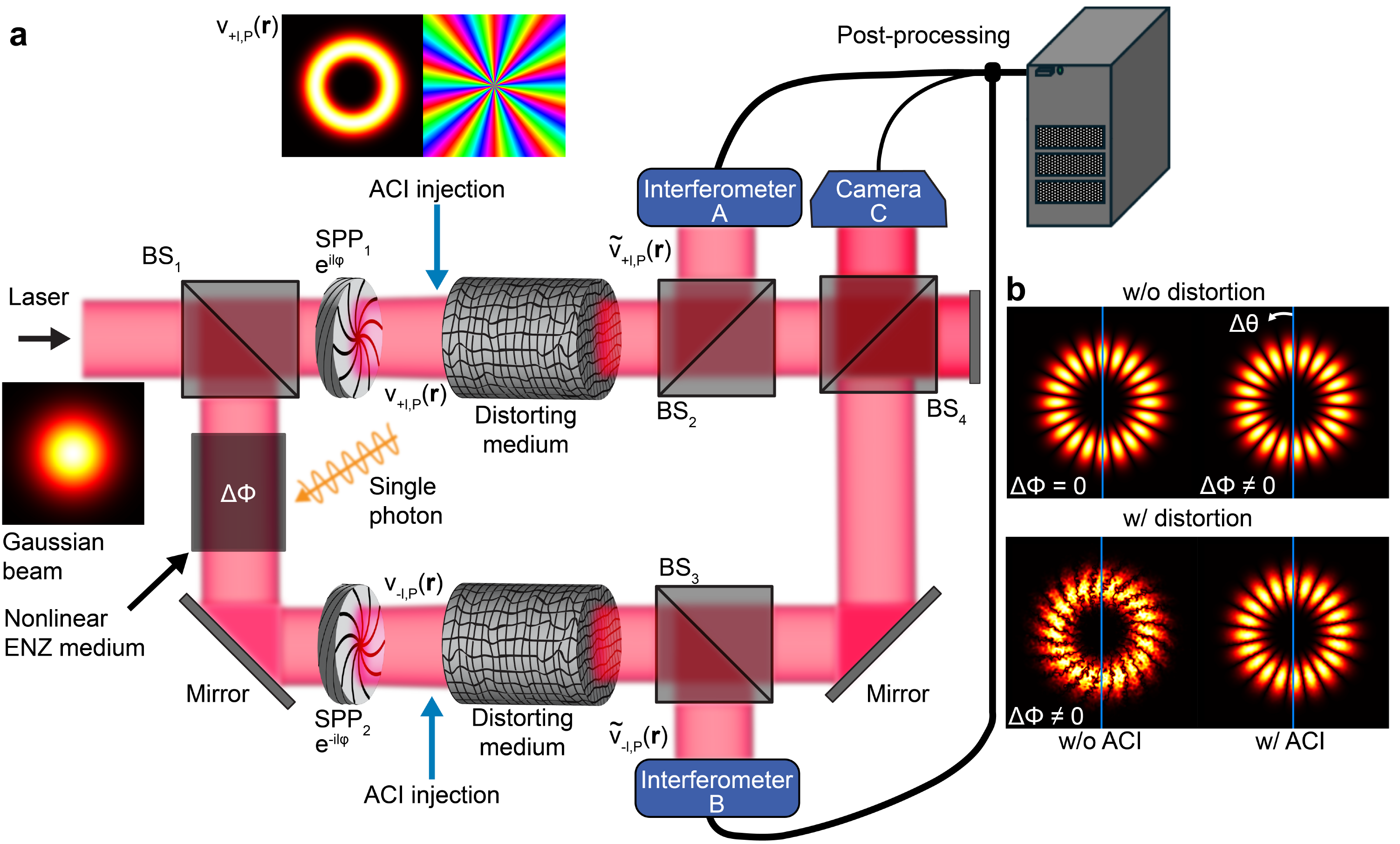}
\caption{(a) Conceptual ACI-assisted OAM interferometer for high-precision phase measurement in the presence of propagation-induced distortion. A Gaussian beam is divided at $\mathrm{BS}_1$ into reference and sensing arms. A small relative phase shift $\Delta\Phi$ is assumed to be introduced in the sensing arm due to, for example, the interaction of single photons with a strongly nonlinear Kerr medium such as an epsilon-near-zero (ENZ) material. Spiral phase plates $\mathrm{SPP}_1$ and $\mathrm{SPP}_2$ produce OAM beams, $v_{+l,P}(\mathbf{r})$ and $v_{-l,P}(\mathbf{r})$, with opposite topological charges $\pm l$. Following propagation through the distorting media, $\mathrm{BS}_2$ and $\mathrm{BS}_3$ direct portions of the fields to detection ports A and B. The complex fields, $\tilde{v}_{+l,P}(\mathbf{r})$ and $\tilde{v}_{-l,P}(\mathbf{r})$, could be recovered using the interferometric arrangement in Fig.~\ref{fig:DemoACIPhaseExrtraction} and used to estimate the corresponding CISs. During subsequent measurements, the CISs would be injected through spatial-domain convolution at the indicated locations before propagation through the distorting media. The remaining beam portions are recombined at $\mathrm{BS}_4$, and camera C records the resulting interference pattern. (b) Representative petal patterns illustrating phase-to-rotation readout using Laguerre-Gaussian modes. Without distortion, single photon excitation of the ENZ material produces an angular displacement $\Delta\theta$ relative to the unexcited reference. Propagation-induced distortion degrades the fringe structure and obscures the rotation. The ACI-assisted pattern illustrates the intended restoration of fringe visibility and petal structure, supporting recovery of the small $\Delta\Phi$ provided that the compensation preserves the excitation-induced relative phase. Two interferometers A and B are used if the distorting media are different.}
\label{fig:OAM_Interferometry}    
\end{figure*}

Without ACI, the reconstructed $36\times36$ beam-array frame in Fig.~\ref{fig:DemoACIFSO}(b) exhibits pronounced spatial deformation, blurring, and loss of contrast, making occupied and unoccupied grid sites difficult to distinguish. The corresponding NCC is $\xi_{P}=0.516$. Similar degradation is observed across the $111$ beam-array frames, producing numerous decoding errors and substantial corruption of the recovered bit sequences. Reassembling these sequences yields the severely degraded grayscale image in Fig.~\ref{fig:DemoACIFSO}(e), in which the original portrait is largely unrecognizable. With ACI, the reconstructed frame in Fig.~\ref{fig:DemoACIFSO}(c) retains distinguishable individual beams and their spatial arrangement, with strong contrast and limited residual distortion. The improved agreement with the reference frame is reflected in the higher NCC of $\xi_{A}=0.955$. Preservation of the array structure facilitates more reliable discrimination between binary states, resulting in substantially fewer decoding errors. After the decoded sequences from all $111$ frames are reassembled, the grayscale image in Fig.~\ref{fig:DemoACIFSO}(f) closely reproduces the reference portrait in Fig.~\ref{fig:DemoACIFSO}(d), although residual pixel-level artifacts remain.

Together, these results motivate further exploration of symbolic spatial encoding, in which multiple bits are represented by the geometry of individual optical symbols. Assigning information to distinguishable spatial patterns offers a compelling alternative to increasing the density of binary beam grids or multiplexed OAM modes. A representative symbol array is illustrated in Fig.~\ref{fig:DemoACIFSO}(g). The symbols are constructed from a base set of 256 Unicode characters by applying a controlled sequence of ten parameterized, noncommuting geometric transformations. The construction is designed to provide a nominal symbol set of size $N_{Sym}=2^{36}$, corresponding to 36 bits per symbol. The $6\times6$ symbol array therefore represents a nominal payload of $36 \times 6 \times 6=1296$ bits, matching that of the beam-array frame in Fig.~\ref{fig:DemoACIFSO}(a). Crucially, for the same nominal 1296-bit payload, the symbolic encoding lowers the spatial resolution demand to $\Delta L_T = 4.34$ (compared to $\Delta L_T = 6.56$ for the binary grid) at the same anisoplanatic ratio $\theta_T = 1.12$. 

The reconstructed intensity distributions obtained without and with ACI are shown in Figs.~\ref{fig:DemoACIFSO}(h) and (i), respectively. Without ACI, turbulence obscures many of the geometric features needed for symbol identification. With ACI, the overall symbol geometry and local features, including loops, tails, and curved strokes, remain more clearly resolved, with the corresponding NCC increasing from $\xi_P=0.428$ to $\xi_A=0.938$. Decoding in such a scheme would rely on pattern recognition, making the geometric distinguishability of the received symbols the central requirement for recovery, without requiring an orthogonal modal basis.

As a second representative application, we examine how ACI could be integrated into orbital angular momentum (OAM) interferometry for potential high-precision phase measurement applications. In the modified Mach-Zehnder setup shown in Fig.~\ref{fig:OAM_Interferometry}, an input Gaussian beam is split by beam splitter $\mathrm{BS}_1$ into an upper reference arm and a lower sensing arm containing a phase-modulating sample. For the illustrative example, the sample is assumed to be a nonlinear epsilon-near-zero (ENZ) material. Due to its strongly nonlinear response \cite{alam2018large,reshef2017beyond,dal2025nonlinear,negro2026mid,kashef2026path,poddar2026effective}, single-photon excitation is assumed to induce a small additional phase shift $\Delta\Phi$ in the sensing beam relative to propagation without single-photon interaction. Two spiral phase plates ($\mathrm{SPP}_1$ and $\mathrm{SPP}_2$) impart opposite helical phase profiles to the reference and sensing beams, producing OAM modes $v_{+l,P}(\mathbf{r})$ and $v_{-l,P}(\mathbf{r})$ with topological charges $+l$ and $-l$, respectively. For the present example, the reference and sensing OAM beams are approximated by Laguerre-Gaussian modes \cite{sueda2004laguerre}. 

After coherent superposition of the two OAM modes at $\mathrm{BS}_4$, an azimuthal interference pattern comprising $2l$ radially oriented intensity lobes (petals) is generated and recorded at camera C. Two sets of petal patterns corresponding to the unexcited and excited states of the ENZ material may be obtained at camera C. In the absence of distortion, representative intensity distributions for $\Delta\Phi=0$ and $\Delta\Phi\neq 0$ are illustrated in Fig.~\ref{fig:OAM_Interferometry}(b). When $\Delta\Phi\neq 0$, the differential phase acquired by the sensing OAM mode produces an angular displacement $\Delta\theta$ of the petal pattern relative to the unexcited state, given by
\begin{equation}
\Delta\theta = \frac{\Delta\Phi}{2l}.
\label{eq:PhaseToRotation}
\end{equation}
This phase-to-rotation transduction enables phase changes to be inferred directly from angular displacement tracking of the measured intensity pattern \cite{kerschbaumer2022twisted}. The achievable phase resolution depends on the precision of angular tracking and the selected topological charge. However, spatial phase fluctuations introduced by a random medium along either or both propagation paths can perturb the beam wavefronts, deforming the multi-petal fringe structure into an irregular speckle pattern under sufficiently strong distortion. Measurement noise at camera C can further obscure the angular displacement, potentially preventing reliable estimation of $\Delta\theta$, particularly when $\Delta\Phi$ is small.

Applying ACI to the reference and sensing beams before propagation through the distorting medium could help preserve the petal fringe structure and improve the reliability of $\Delta\theta$ estimation. The OAM modes $v_{+l,P}(\mathbf{r})$ and $v_{-l,P}(\mathbf{r})$ may serve as probes for characterizing distortion in the reference and sensing arms, respectively. As illustrated in Fig.~\ref{fig:OAM_Interferometry}, two additional beam splitters, $\mathrm{BS}_2$ and $\mathrm{BS}_3$, divert portions of the reference and sensing beams to interferometers A and B, respectively. The interferometers are implemented as in Fig.~\ref{fig:DemoACIPhaseExrtraction} to recover the reconstructed full complex probe fields, $\tilde{v}_{+l,P}(\mathbf{r})$ and $\tilde{v}_{-l,P}(\mathbf{r})$, from intensity measurements, before the beams are superimposed at $\mathrm{BS}_4$, enabling estimation of the corresponding CISs using Eq.~\ref{eq:CISMapping}. 

The estimated CISs for the reference and sensing beams may then be injected into $v_{+l,P}(\mathbf{r})$ and $v_{-l,P}(\mathbf{r})$, respectively, through spatial-domain convolution before propagation through the distorting medium. The proposed injection locations are indicated in Fig.~\ref{fig:OAM_Interferometry}. Since the corresponding input planes are directly accessible, injection could be implemented using two additional SLMs, one in each arm, following the concept illustrated in  Fig.~\ref{fig:ACIFullSchematic_partB}. As illustrated schematically in the lower-right portion of Fig.~\ref{fig:OAM_Interferometry}(b), accurately estimated and implemented CISs could help restore fringe visibility and preserve the petal structure under strongly distorting conditions. Provided that the compensation also preserves the excitation-induced relative phase, the restored pattern could support reliable estimation of $\Delta\theta$ and recovery of $\Delta\Phi$. Therefore, future work should focus on the quantitative analysis of the minimum detectable phase shift with ACI.     

Related extensions could be explored in biomedical imaging, where variations in cellular thickness and refractive index introduce optical-path-length differences, while tissue scattering can obscure the associated interference patterns. Combining ACI distortion compensation with OAM interferometric readout and suitable spatially resolved phase reconstruction could provide a potential route toward non-invasive, label-free quantitative phase imaging (QPI) in turbid biological media \cite{aknoun2021quantitative,farjaminejad2025non}. Recovering weak phase signatures associated with nanoscale optical-path-length variations at greater tissue depths represents a prospective research objective.

\section{Conclusion}

We have presented a unified review of the ACI framework for preserving highly structured optical fields during propagation through lossy and noisy systems and dynamic random media. By combining segmented reciprocal-space characterization with spectrally localized compensation injection, ACI aims to extend reliable distortion mitigation into propagation regimes where conventional correction techniques can become unstable or ill-conditioned. We have explored potential applications of ACI and identified practical deployment considerations, including optical access, measurement fidelity, calibration transfer, and the time required for probe acquisition, CIS estimation, and compensation updates relative to the channel decorrelation time. Further work should focus on rigorous experimental validation of the proposed architectures and their adaptation to specific operating conditions. Potential applications include deep-tissue imaging and tomography, optogenetics, FSO communications, remote sensing, photon number resolving and quantum non-demolition single-photon detection. Across these fields, the broader opportunity is to improve the precision of light delivery and increase the complexity of optical information that can be reliably transmitted and recovered despite cumulative, spatiotemporally varying distortions.

\section*{Acknowledgments}

We thank Seth Nelson for useful discussions on the OAM based sensing. This work was supported in parts through the QEMISD MURI project funded by ARO-W911NF-24-2-0195 P00001; ARO W911NF 2510283; and by the National Science Foundation (Grant No. EAR-2221730).

\section*{Data Availability}

The data that support the findings of this study are available on reasonable request.

%\bibliography{references}% Produces the bibliography via BibTeX.
%apsrev4-2.bst 2019-01-14 (MD) hand-edited version of apsrev4-1.bst
%Control: key (0)
%Control: author (8) initials jnrlst
%Control: editor formatted (1) identically to author
%Control: production of article title (0) allowed
%Control: page (0) single
%Control: year (1) truncated
%Control: production of eprint (0) enabled
%

\end{document}